\documentclass[a4paper,11pt]{article}%
\usepackage{amsmath,xcolor}
\usepackage{graphicx}
\usepackage{amsbsy}
\usepackage{cite}
\usepackage{slashed}
\usepackage{epsfig}
\usepackage{amsfonts}
\usepackage{amssymb}
\usepackage[T1]{fontenc}
\usepackage[utf8]{inputenc}
\usepackage{lmodern}
\newcommand{\bm}[1]{\boldsymbol{#1}}
\begin{document}
\thispagestyle{empty} \setcounter{page}{0}
\begin{flushright}
CP3-13-27\\
LPSC13158\\
July 2013\\
\end{flushright}

\vskip 2.1 true cm

\begin{center}
{\huge The same-sign top signature of R-parity violation}\\[1.9cm]

\textsc{Gauthier Durieux}$^{1}$ \textsc{and Christopher Smith}$^{2}$
\vspace{0.5cm}
\\[9pt]
$^{1}\;${\small\textsl{Centre for Cosmology, Particle Physics and Phenomenology (CP3),}}\newline 
{\small \textsl{Universit\'{e} catholique de Louvain, Chemin du Cyclotron 2, B-1348 Louvain-la-Neuve, Belgium}}
\\\smallskip
$^{2}\;${\small\textsl{LPSC, Universit\'{e} Joseph Fourier Grenoble 1, CNRS/IN2P3 UMR5821, }}\newline {\small \textsl{Institut Polytechnique de Grenoble, 53 rue des Martyrs, 38026 Grenoble Cedex, France.}}
\\[1.9cm]
\textbf{Abstract}\smallskip
\end{center}

\begin{quote}
\noindent 
Baryonic R-parity violation could explain why low-scale supersymmetry has not yet been discovered at colliders: sparticles would be hidden in the intense hadronic activity. However, if the known flavor structures are any guide, the largest baryon number violating couplings are those involving the top/stop, so a copious production of same-sign top-quark pairs is in principle possible. Such a signal, with its low irreducible background and efficient identification through same-sign dileptons, provides us with tell-tale signs of baryon number violating supersymmetry. Interestingly, this statement is mostly independent of the details of the supersymmetric mass spectrum. So, in this paper, after analyzing the sparticle decay chains and lifetimes, we formulate a simplified benchmark strategy that covers most supersymmetric scenarios. We then use this information to interpret the same-sign dilepton searches of CMS, draw approximate bounds on the gluino and squark masses, and extrapolate the reach of the future $14$~TeV runs.

\let \thefootnote \relax
\footnotetext{$^{1}\;$gauthier.durieux@uclouvain.be}
\footnotetext{$^{2}\;$chsmith@lpsc.in2p3.fr} 
\end{quote}
\newpage

%%%%%%%%%%%%%%%%%%%%%%%%%%%%%%%%%%%%%%%%%%%%%%%%%%%%%%%%%%%%%%%%%%%%
\section{Introduction}

After two years of operation, the LHC experiments have not found any signal of low-scale supersymmetry. Current mass bounds on simple supersymmetric scenarios are now pushed beyond the TeV. This is especially striking in the simplified setting where squarks, gluino, and neutralinos are the lightest supersymmetric degrees of freedom. With the gluino and all the squarks degenerate in mass, the bounds are above $1.5$ TeV~\cite{AtlasSUSYReach,CMSSUSYReach}.

Most searches for supersymmetry are done assuming R parity is exact, thereby forbidding all baryon number violating (BNV) and lepton number violating (LNV) couplings~\cite{Barbier04}. Indeed, at first sight, the incredibly tight limits on the proton decay lifetime~\cite{PDG} seem to lead to an unacceptable fine-tuning of these couplings. But, imposing R parity is not innocuous for the phenomenology of the Minimal Supersymmetric Standard Model (MSSM). Most dramatically, superpartners have to be produced in pairs and the lightest supersymmetric particle (LSP) is absolutely stable. It thus contributes to the dark matter density of the Universe, and has to be electrically neutral and colorless. So, at colliders, all superpartners cascade decay down to this LSP, which manifests itself as missing energy. In particular, the tight bounds on the gluino and the squark masses are derived looking for the signatures of such cascade decays down to the invisible LSP.

The hypothesis of an exact R parity is thus entwined within current search strategies. This predicament mostly remains even though the original motivation for R parity no longer holds. As was shown in Ref.~\cite{RPVMFV}, the BNV and LNV couplings do not require any fine tuning to comply with the proton decay bounds. Rather, being flavored couplings, they just need to be aligned with the flavor structures already present in the Standard Model (SM). In this way, the strong hierarchies of the known fermion masses and mixings, e.g. $m_{\nu}\ll m_{u}\ll m_{t}$, are passed on to the R-parity violating (RPV) couplings. Consequently, low-energy observables, mainly sensitive to the very suppressed first-generation RPV couplings, naturally comply with all existing bounds.

\subsection{Theoretical framework}

To precisely define and enforce the alignment of the RPV couplings with the SM flavor structures, the Minimal Flavor Violation (MFV) framework is ideally suited~\cite{MFV}. This is the approach proposed in Ref.~\cite{RPVMFV}, of which we only sketch the main line of arguments here. The starting point of the MFV hypothesis is the assumption that, at least in a first approximation, the Yukawa couplings $\mathbf{Y}_{u}$, $\mathbf{Y}_{d}$, and $\mathbf{Y}_{e}$ are the only explicit breaking terms (or spurions) of the $SU(3)^{5}$ flavor symmetry exhibited by the MSSM gauge interactions. Then, all the other flavor couplings, including those violating R parity, are constructed out of these spurions in a manifestly $SU(3)^{5}$ invariant way. The main result of this analysis is that the transformation properties of the Yukawa couplings under $SU(3)^{5}$ allow only for the BNV couplings,%
\begin{equation}
W_{RPV}= \bm{\lambda}^{\prime\prime IJK}U^{I}D^{J}D^{K}\;,\label{UDD1}%
\end{equation}
where $I,J,K$ are flavor indices. Specifically, MFV leads to expressions like%
\begin{equation}
\bm{\lambda}^{\prime\prime IJK}=\varepsilon_{LMN}\mathbf{Y}_{u}^{IL}\mathbf{Y}_{d}^{JM}\mathbf{Y}_{d}^{KN}\oplus\varepsilon_{LJK}(\mathbf{Y}_{u}\mathbf{Y}_{d}^{\dagger})^{IL}\oplus\varepsilon_{IMN}(\mathbf{Y}_{d}\mathbf{Y}_{u}^{\dagger})^{JM}(\mathbf{Y}_{d}\mathbf{Y}_{u}^{\dagger})^{KN}\oplus...\;,\label{UDD2}%
\end{equation}
where $\oplus$ serves as a reminder that arbitrary order one coefficients are understood for each term. By contrast, none of the LNV couplings can be constructed out of the leptonic Yukawa coupling $\mathbf{Y}_{e}$. Even introducing a neutrino Dirac mass term does not help. Actually, it is only once a left-handed neutrino Majorana mass term is included in the spurion list that such couplings can be constructed, but they are then so tiny that they are completely irrelevant for collider phenomenology.

Obviously, once this alignment hypothesis is enforced, the $\bm{\lambda}^{\prime\prime}$ couplings are highly hierarchical. However, the predicted hierarchy depends on additional parameters or assumptions besides MFV itself. First, they strongly depend on $\tan\beta=v_{u}/v_{d}$, the ratio of the vacuum expectation values of the two MSSM neutral Higgs bosons, since $\mathbf{Y}_{d}\ll\mathbf{Y}_{u}$ when $\tan\beta\lesssim5$ (see Table~\ref{Hierarchies}). Then, specific models might not generate all the possible structures shown in Eq.~(\ref{UDD2}). In particular, the holomorphic restriction introduced in Ref.~\cite{Grossman} allows for the first term only,\footnote{For recent discussions of possible dynamical origins for this holomorphic MFV prescription, see Refs.~\cite{Gauged}.} and further forbids introducing flavor-octet combinations like $\mathbf{Y}_{u}^{\dagger}\mathbf{Y}_{u}$ and $\mathbf{Y}_{d}^{\dagger}\mathbf{Y}_{d}$. This last restriction is not RGE invariant though~\cite{RPVRGE}. If the dynamics at the origin of the flavor structures take place at some very high scale, we need to relax the holomorphic constraint. Further, from an effective point of view, such $\mathbf{Y}_{u}^{\dagger}\mathbf{Y}_{u}$ and $\mathbf{Y}_{d}^{\dagger}\mathbf{Y}_{d}$ insertions occur at the low scale through electroweak corrections. So, in the following, we will denote by holomorphic the slightly loser hierarchy derived starting with $\varepsilon_{LMN}\mathbf{Y}_{u}^{IL}\mathbf{Y}_{d}^{JM}\mathbf{Y}_{d}^{KN}$, but allowing for additional non-holomorphic $\mathbf{Y}_{u}^{\dagger}\mathbf{Y}_{u}$ and $\mathbf{Y}_{d}^{\dagger}\mathbf{Y}_{d}$ spurion insertions (see Table~\ref{Hierarchies}).%

\begin{table}[t] \centering
$%
\begin{tabular}[c]{ccc}\hline
$\bm{\lambda}_{IJK}^{\prime\prime}$ & Full MFV & Holomorphic MFV\\\hline
& $\;\;\;\;\;\begin{array}[c]{ccc} ds\;\; & \;\;sb\;\; & \;\;db \end{array}$ & 
$ \begin{array}[c]{ccc} ds\;\; & \;\;sb\;\; & \;\;db \end{array}$\\
$\tan\beta=5$ & $\begin{array}[c]{c} u\\ c\\ t \end{array}
\left( \begin{array}[c]{ccc}
10^{-5} & 10^{-5} & 10^{-5}\\
10^{-4} & 10^{-6} & 10^{-5}\\
0.1 & 10^{-5} & 10^{-4}
\end{array} \right)  $ & 
$\left( \begin{array}[c]{ccc}
10^{-13} & 10^{-8} & 10^{-10}\\
10^{-10} & 10^{-6} & 10^{-7}\\
10^{-6} & 10^{-5} & 10^{-6}
\end{array} \right)  $\\
$\tan\beta=50$ & $\begin{array}[c]{c} u\\ c\\ t \end{array}
\left( \begin{array}[c]{ccc}
10^{-4} & 10^{-4} & 10^{-4}\\
10^{-3} & 10^{-4} & 10^{-4}\\
1 & 10^{-3} & 10^{-3}
\end{array} \right)  $ & 
$\left( \begin{array}[c]{ccc}
10^{-11} & 10^{-6} & 10^{-9}\\
10^{-8} & 10^{-4} & 10^{-5}\\
10^{-4} & 10^{-3} & 10^{-4}
\end{array} \right)  $\\\hline
\end{tabular}
\ \ $
\caption{Hierarchies predicted for the $\Delta B=1$ R-parity violating coupling, under the full MFV hypothesis~\cite{RPVMFV} and under its holomorphic restriction~\cite{Grossman}. In this latter case, we adopt a slightly looser definition to account for possible RGE effects and to stabilize the hierarchies under electroweak corrections (see the discussion in the main text; all these numbers are taken from Ref.~\cite{RPVMFV}). Because $\bm{\lambda}^{\prime\prime}_{IJK}$ is antisymmetric under $J\leftrightarrow K$, its entries can be put in a $3\times3$ matrix form with $I = {u,c,t}$ and $JK = {ds,sb,db}$.}%
\label{Hierarchies}%
\end{table}%

It is clear from Table~\ref{Hierarchies} that no matter the precise MFV implementation, the largest BNV couplings are always those involving the top (s)quark. Those with up or charm (s)quarks are extremely small, essentially because the epsilon tensor antisymmetry forces them to be proportional to light-quark mass factors (see Eq.~(\ref{UDD2})). This permits to satisfy all the low energy constraints from proton decay or neutron oscillations, even for squark masses below the TeV scale. In this context, it is worth to stress that by construction, the MFV hierarchies are stable against electroweak corrections. So, the MFV implementation can be interpreted as a way to maximize the $\bm{\lambda}_{IJK}^{\prime\prime}$ coupling for each $I,J,K$. For example, if $\bm{\lambda}_{tds}^{\prime\prime}$ exceeds the value shown in Table~\ref{Hierarchies}, it may induce a larger effective $\bm{\lambda}_{uds}^{\prime\prime}$ coupling through SM or MSSM flavor transitions, and thereby conflict with experimental constraints.

\subsection{Search strategy at colliders}

The presence of the RPV couplings deeply alters the supersymmetric collider phenomenology, and none of the sparticle mass bounds set in the R-parity conserving case are expected to survive. So, it is our purpose here to analyze the signatures of the MSSM supplemented with the $UDD$ coupling of Eq.~(\ref{UDD2}), under the assumptions that $\bm{\lambda}^{\prime\prime}$ follows the hierarchies shown in Table~\ref{Hierarchies}. Before entering the core of the discussion, let us expose our strategy.

Since low energy constraints allow some of the BNV couplings to remain relatively large, no supersymmetric particle is expected to live for long. Except in a small corner of parameter space (to be detailed later), only SM particles are seen at colliders. The simplest amplitudes with intermediate (on-shell or off-shell) sparticles are thus quadratic in the BNV couplings, and correspond either to $\Delta B=0$ or $\Delta B=\pm2$ transitions. Typically, the former takes the form of enhancements in processes with SM-allowed final states, like $t\bar{t}+\text{jets}$ or multijet processes. Except if a resonance can be spotted, these are rather non-specific signatures, and one must fight against large backgrounds. On the other hand, the $\Delta B=\pm2$ channels have much cleaner signatures which, to a large extent, transcend the details of the MSSM mass spectrum. Indeed, regardless of the underlying dynamics, the MFV hierarchy strongly favors the presence of two same-sign top quarks in the final state. A sizable same-sign lepton production is therefore predicted. At the same time, the initial state at the LHC has a $B=+2$ charge since it is made of two protons. As analyzed model-independently in Ref.~\cite{DurieuxGMS12}, this can induce a significant negative lepton charge asymmetry, which is defined as
\begin{equation}
A_{\ell\ell^{\prime}}\equiv\frac{\sigma(pp\rightarrow\ell^{+}\ell^{\prime}%
{}^{+}+X)-\sigma(pp\rightarrow\ell^{-}\ell^{\prime}{}^{-}+X^{\prime})}%
{\sigma(pp\rightarrow\ell^{+}\ell^{\prime}{}^{+}+X)+\sigma(pp\rightarrow
\ell^{-}\ell^{\prime}{}^{-}+X^{\prime})}\;.
\label{Asym}%
\end{equation}
Observing $A_{\ell\ell'}<0$ would not only point clearly at new physics, but also strongly hint at baryon number violation. Indeed, the SM as well as most new physics scenarios generate positive asymmetries.

In the present paper, we will thus concentrate on this same-sign dilepton signal and its associated charge asymmetry. The other prominent RPV signatures, namely multijet resonances and R-hadron states, are described in the next section. To quantify the relative strengths of these signatures, it is necessary to analyze in some details the various mass hierarchies and decay chains. This is done in section 3, where the most relevant degrees of freedom are identified (the calculations of the squark, gluino, and neutralino decay rates in the presence of the $\bm{\lambda}^{\prime\prime}$ couplings are briefly reviewed in appendix~\ref{AppDecay}). We then show in section 4 how this information permits to set up a simplified framework. In section 5, we use this benchmark to translate the current experimental limits into constraints on sparticle masses, and to analyze the sensitivity of the future $14$~TeV runs. Finally, our results are summarized in the conclusion.

%%%%%%%%%%%%%%%%%%%%%%%%%%%%%%%%%%%%%%%%%%%%%%%%%%%%%%%%%%%%%%%%%%%%
\section{Characteristic signatures of the R-parity violating MSSM}

In the R-parity conserving case, the simplest production mechanisms for supersymmetric particles at the LHC are driven by the supersymmetrized QCD part of the MSSM. Further, processes like $d\,d\rightarrow\tilde{d}\,\tilde{d}\,$ or $\,g\,g\rightarrow\tilde{g}\,\tilde{g}$ have very large cross-section when the on-shell $\tilde{d}$ or $\tilde{g}$ production is kinematically accessible, hence the tight bounds already set on these particle masses. As stressed in the introduction, these bounds assume the presence of a significant missing energy in the final state and only hold if R parity is conserved.

\begin{figure}[t]
\centering
\includegraphics[width=15.5cm]{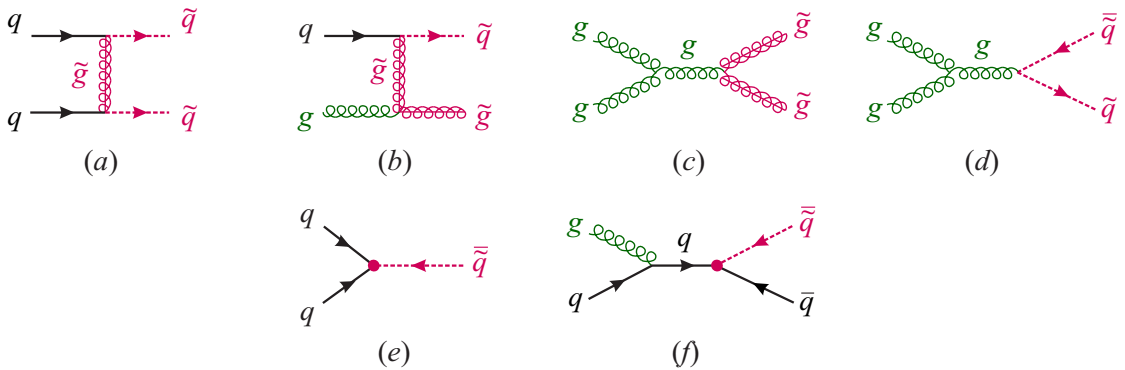}
\caption{Some dominant leading-order strong ($a-d$) and RPV ($e$) production mechanisms of squarks and gluinos at the LHC. Processes with initial gluons or proton valence quarks, $q=u,d$, are favored by the parton distribution functions. We also show the next-to-leading order resonant squark production mechanism ($f$) because the dominant RPV couplings, shown as red dots, involve either the $t,d,s$ flavors in the full MFV case, or $t,d,b$; $t,d,s$; and $t,s,b$ flavors in the holomorphic MFV case, and thus the diagram ($e$) necessarily involves at least one sea quark.}%
\label{Production}%
\end{figure}

\begin{figure}[t]
\centering \includegraphics[width=16.5cm]{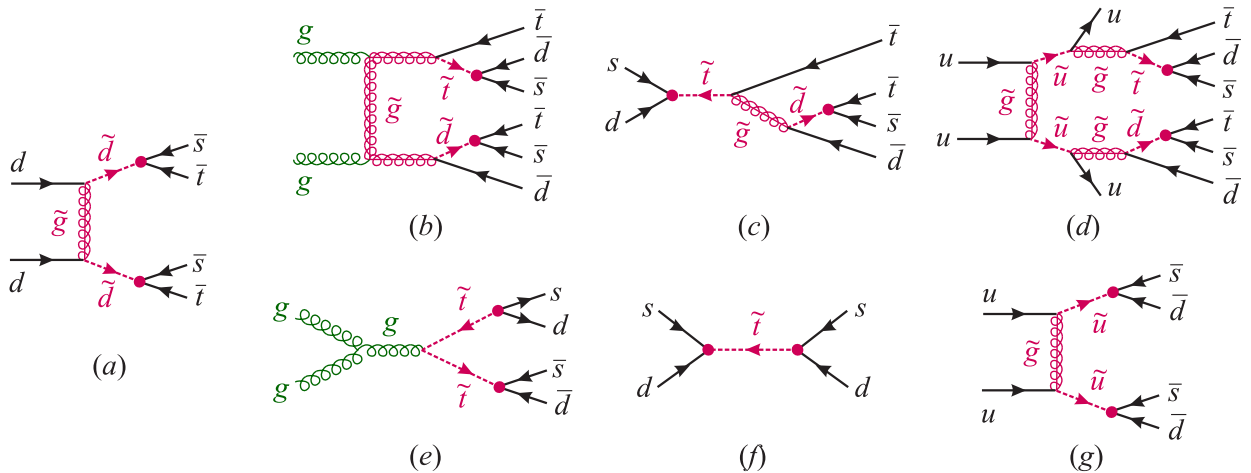}
\caption{($a-d$) Examples of mechanisms leading to same-sign top pair final states, starting from the dominant QCD processes of Fig.~\ref{Production}. ($e-g$) Examples of production mechanisms leading to light-quark jet final states.}%
\label{Signals}%
\end{figure}

When the largest RPV coupling is smaller or comparable to $\alpha_{S}$, squarks and gluinos are still mostly produced in pair through QCD processes. The main non-QCD mechanism yielding sparticles is the single squark resonant production, which requires less center-of-mass energy. At the LHC, the most abundantly produced sparticle states are thus (considering for now the full MFV hierarchy, see Table~\ref{Hierarchies}):
\begin{equation}
\begin{aligned}
u\,u  &	\rightarrow\tilde{u}_{L,R}\;\tilde{u}_{L,R}\;,\quad
d\,d	\rightarrow\tilde{d}_{L,R}\;\tilde{d}_{L,R}\;,\quad
u\,d	\rightarrow\tilde{u}_{L,R}\;\tilde{d}_{L,R}\;,\\
g\,g  &	\rightarrow\tilde{g}\,\tilde{g}\;,\quad
g\,g	\rightarrow\tilde{q}_{L,R}\;\overline{\tilde{q}}_{L,R}\;,\\
s\,d  &	\rightarrow\overline{\tilde{t}}_{R}\;,
\end{aligned}
\label{LHCProd}%
\end{equation}
and are shown in Fig.~\ref{Production}. The main difference with the R-parity conserving case is that once the $\bm{\lambda}^{\prime\prime}$ couplings are turned on, each of these sparticles initiates a decay chain ending with quark final states, resulting in a significant hadronic activity instead of missing energy. If we assume that the charginos and sleptons are heavier than squarks, gluinos, and the lightest neutralino (denoted simply as $\tilde{\chi}^{0}\equiv\tilde{\chi}_{1}^{0}$ in the following), then we can identify three main characteristic signatures in this hadronic activity:
\begin{enumerate}
\item \textbf{Top-quark production} including same-sign top pairs. Because the dominant $\bm{\lambda}_{IJK}^{\prime\prime}$ couplings are those with $I=3$, most processes lead to top quarks in the final states (see Fig.~\ref{Signals}). For example, we have $\tilde{d}\rightarrow\bar{t}\,\bar{s}\:$ or $\:\tilde{g},\tilde{\chi}^{0}\rightarrow t\,d\,s,\;\bar{t}\,\bar{d}\,\bar{s}$. Even the stop can decay into top-quark pairs if $\tilde{t}\rightarrow\tilde{g}\,t$ or $\tilde{t}\rightarrow\tilde{\chi}^{0}\,t$ is kinematically open (see Fig.~\ref{Signals}$c$). For all these modes, a crucial observation is that the production of same-sign top pairs is always possible thanks to the Majorana nature of the gluino and neutralino. Despite its relatively small $5\%$ probability, the same-sign dilepton signature is best suited for identifying such final states. There are several reasons for this. First, charged leptons are clearly identified in detectors and avoid jet combinatorial background. Second, they allow to determine almost unambiguously the sign, and therefore the baryon number, of the top quarks they arose from. Finally, irreducible backgrounds are small as same-sign dilepton production is rare in the SM. So, this is the signature on which we will concentrate in the following (see also Refs.~\cite{Allanach:2012vj,AsanoRS13,BergerPST13}).

\item \textbf{Di- or trijet resonances} built over light quarks and maybe a few $b$ quarks. A priori, dijets could originate from squark decays and trijets from gluino or neutralino decays. But with MFV, only up-type intermediate squarks can lead to light-quark jets, since the other sparticle decay products always include a top quark. The simplest process is thus the $\Delta B=0$ resonant stop production with a dijet final state (see Fig.~\ref{Signals}$f$). But since the electric charge of a jet is not measurable, the $\Delta B$ nature of the transition cannot be ascertained and QCD backgrounds appear overwhelming. Nevertheless, given the potentially large cross sections of the strong production processes depicted in Fig.~\ref{Production}, such an enhanced jet activity could be accessible experimentally~\cite{Stops}, and has already been searched for at colliders (see e.g. Ref.~\cite{ThreeJets}).

\item \textbf{Long-lived exotic states}, the so-called R-hadrons built as hadronized squarks or gluinos flying away~\cite{Rhadrons}. Such quasi-stable exotic states have already been searched for experimentally, excluding squark masses below about $600$~GeV and gluino masses below about $1$~TeV~\cite{StablesQ}. But, as will be detailed in the next section, R-hadron signatures are rather difficult to get once MFV is imposed. Indeed, some RPV couplings are large and all sparticles can find a way to use them for decaying. For example, if $\bm{\lambda}_{tds}^{\prime\prime}\approx0.1$, then $\tilde{g},\tilde{\chi}^{0}\rightarrow t\,d\,s,\;\bar{t}\,\bar{d}\,\bar{s}$ proceeding via a virtual squark or $\tilde{q}_{L,R}\rightarrow q\,t\,d\,s,\;q\,\bar{t}\,\bar{d}\,\bar{s}$ mediated by a virtual gaugino and a virtual squark (see Fig.~\ref{Signals}$d$) are kinematically available and occur rather quickly for masses below $1$~TeV (this is also true for a slepton LSP, see appendix~\ref{App4B}). Note, however, that very large gluino (or neutralino) lifetimes can always be obtained by increasing the squark masses well beyond the TeV scale, as for example in the split SUSY scenario~\cite{SplitSUSY}. 
\end{enumerate}

The relative and absolute strengths of these signals depend crucially on the MSSM mass spectrum. To proceed, we analyze in the next section the different spectra and corresponding decay chains in some details. This is a rather technical discussion, further complemented by the decay rate calculations in appendix A, whose main outcomes are depicted in Figs.~\ref{DecaysFig} and~\ref{FigLongLife}. The former shows that most sparticle decay chains end with top quarks, while the latter shows that the LSP lifetimes are nearly always short enough to avoid R-hadron constraints. Provided these two pieces of information are kept in mind, the reader less inclined to go through all the details may wish to directly jump to section 4, where our simplified setting is put in place.

%%%%%%%%%%%%%%%%%%%%%%%%%%%%%%%%%%%%%%%%%%%%%%%%%%%%%%%%%%%%%%%%%%%%
\section{Sparticle decay chains and lifetimes}

\begin{figure}[t]
\centering
\includegraphics[width=13.2cm]{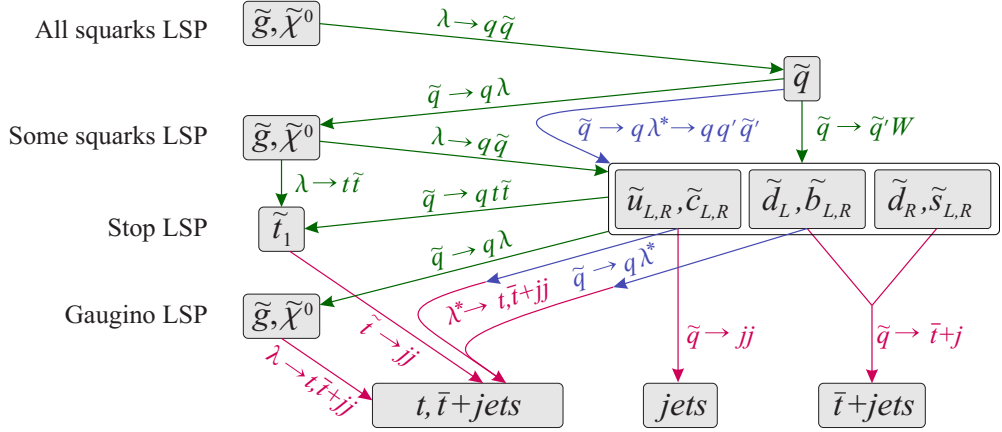}
\caption{Decay chains of the squarks, gluino, and lightest neutralino down to quark-only final states, depending on whether the gauginos, the stop, some of the squarks, or all the squarks are the lightest supersymmetric particles. The symbol $\lambda^{(*)}$ denotes a real (virtual) gluino or neutralino. For each squark, the relative strengths of the R-parity conserving (green and blue) and R-parity violating (red) transitions depend on the details of the mass spectrum as well as on the MFV hierarchy. In particular, whenever the gluino (and to some extent, neutralino) is too heavy to be produced on-shell (green), its virtual exchange opens some decay channels (blue) competing with the direct RPV decay processes (red). In the full MFV hierarchy, where $\boldsymbol{\lambda}_{tds}^{\prime\prime}$ is the largest RPV coupling, the jets arise mostly from $s$ and $d$ quarks. In the holomorphic case, some of them are built upon $b$ quarks instead.}%
\label{DecaysFig}%
\end{figure}

The various possible cascades are depicted in Fig.~\ref{DecaysFig}. With charginos and sleptons decoupled, two alternative cases can be distinguished depending on whether the gluino or the squarks are the lighter.

\subsection{Gluino lighter than squarks}

Let us concentrate first on the lower-left corner of this diagram. Still assuming that QCD processes dominate over RPV ones, the decay chains preferentially start by $\tilde{q}\rightarrow q\,\tilde{g}$ when gluinos are lighter than squarks. These transitions are overwhelmingly flavor conserving when MFV is enforced. If the gluino is the LSP, it then decays through the RPV coupling: $\tilde{g}\rightarrow t\,d\,s,\;\bar{t}\,\bar{d}\,\bar{s}$ (the full MFV hierarchy is assumed for now). If the lightest neutralino is the LSP, it is produced via $\tilde{g}\rightarrow q\,\bar{q}\,\tilde{\chi}^{0}$, $\tilde{q}\rightarrow q\,\tilde{\chi}^{0}$, as well as directly from electroweak processes, and decays again as $\tilde{\chi}^{0}\rightarrow t\,d\,s,\;\bar{t}\,\bar{d}\,\bar{s}$. Along these chains, the branching ratios are all close to $100\%$, except for the electroweak $\tilde{q}_R\rightarrow q\,\tilde{\chi}^0$ with which the fastest direct RPV decays $\tilde{q}_R\rightarrow \bar{q}\,\bar{q}^{\prime}$ could compete.

\begin{figure}[t]
\centering
\includegraphics[width=0.96\textwidth]{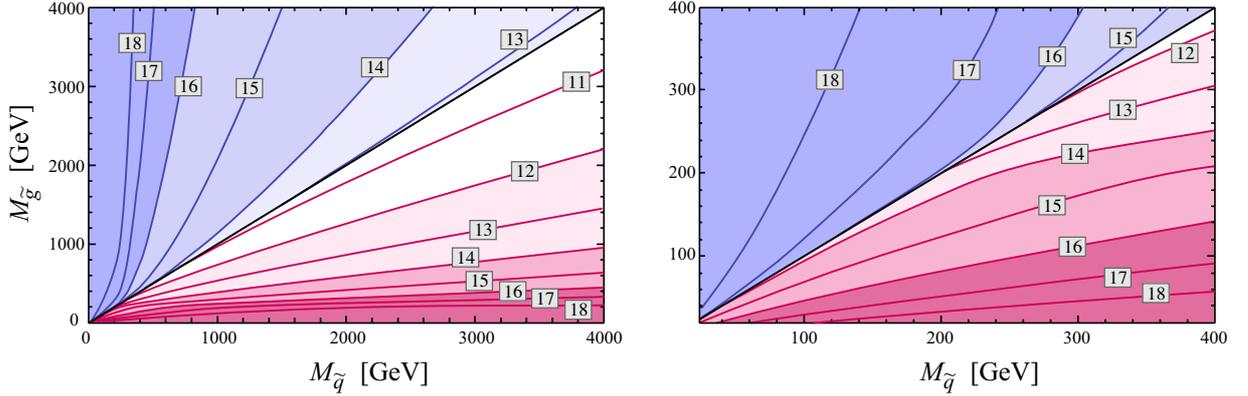}  
\caption{Colored LSP partial widths in the holomorphic MFV case with $\tan\beta = 5$. Labels stand for $-\log_{10}(\Gamma~\text{[GeV]})$. Specifically, the largest RPV couplings $\bm{\lambda}_{3IJ}^{\prime\prime}\lesssim 10^{-5}$ dominate everywhere except in the low mass region where the top channels get kinematically suppressed and the impact of the subdominant RPV couplings $\bm{\lambda}_{2IJ}^{\prime\prime}\lesssim 10^{-6}$ begins to be felt. Below the diagonal, the gluino is the LSP and decays via a virtual squark, while above the diagonal, the plots show the width of the most stable squarks, assuming it decays exclusively through a virtual gluino and a virtual squark. As explained in the text, this requires turning off the left-right squark mixing terms so as to close the decay channel of Eq.~(\ref{uL3}). Phenomenologically, widths below $10^{-16}$~GeV ($10$ ns) can lead to R-hadron signals~\cite{StablesQ}, those below $10^{-14}$~GeV ($0.1$ ns) could render the top identification difficult (because of the required $b$ tagging~\cite{Buchkremer:2012dn}), while values up to a few $10^{-12}$~GeV ($0.001$ ns) could lead to noticeable displaced vertices~\cite{Grossman}. Note that $\max(\bm{\lambda}_{3IJ}^{\prime\prime})$ and $\max(\bm{\lambda}_{2IJ}^{\prime\prime})$ are the smallest in the holomorphic MFV case with $\tan\beta = 5$, but the plots for any other values can easily be inferred since all decay rates are quadratic in $\bm{\lambda}^{\prime\prime}$. For example, all the widths are $3^4\approx 100$ times larger if $\tan\beta = 15$. In the full MFV case, but still at $\tan\beta = 5$, the widths above (below) the top-quark threshold are $10^8$ ($10^4$) times larger, and even observing displaced vertices becomes impossible over most of the parameter space.}%
\label{FigLongLife}%
\end{figure}

Note that the partial widths of the gluino and neutralino are fairly large. Assuming the lightest neutralino is dominantly a bino, taking all squarks degenerate and neglecting $m_{t}/M_{\tilde{g},\tilde{\chi}^{0}}$ as well as higher powers of $M_{\tilde{g},\tilde{\chi}^{0}}/M_{\tilde{q}}$ (see the discussion in appendix~\ref{App3B}), we get
\begin{align}
\Gamma\Big(\tilde{g}  &  \rightarrow t\,d\,s\Big)
	\approx \frac{3\alpha_{S}M_{\tilde{g}}}{512\pi^{2}} \times |\bm{\lambda}_{tds}^{\prime\prime}|^{2} \times \frac{M_{\tilde{g}}^{4}}{M_{\tilde{q}}^{4}} \;,\\
\Gamma\Big(\tilde{\chi}^{0}  &  \rightarrow t\,d\,s\Big)
	\approx \frac{\alpha M_{\tilde{\chi}^{0}} |N_{1B}|^2}{128\pi^{2}\cos^{2}\theta_{W}} \times |\bm{\lambda}_{tds}^{\prime\prime}|^{2} \times \frac{M_{\tilde{\chi}^{0}}^{4}}{M_{\tilde{q}}^{4}}\;.
\end{align}
Numerically, for $M_{\tilde{q}}\approx1$~TeV and $M_{\tilde{g}}\approx M_{\tilde{\chi}^{0}}\approx300$~GeV, these widths are $\Gamma_{\tilde{g}}\approx(10^{-4}$ GeV$)\times|\bm{\lambda}_{tds}^{\prime\prime}|^{2}$ and $\Gamma_{\tilde{\chi}^{0}}\approx(10^{-5}$ GeV$)\times|\bm{\lambda}_{tds}^{\prime\prime}|^{2}$ (when the lightest neutralino is a pure bino, $|N_{1B}|=1$). We do not consider here the split-SUSY scenario~\cite{SplitSUSY} where squarks are much heavier than the gluino or neutralino, so these numbers represent the minimum lifetimes for these particles. They are short enough to circumvent the already tight experimental bounds set on R-hadrons~\cite{StablesQ}. Actually, except at low $\tan\beta$ and with the holomorphic MFV hierarchy (see Fig.~\ref{FigLongLife}), these sparticles even decay too quickly to leave noticeable displaced vertices.\footnote{If the gluino or neutralino are lighter than the top quark, then they decay into three light quarks thanks to subdominant RPV couplings. In the holomorphic case at low $\tan\beta$, the lifetimes could then be sufficiently large to generate R-hadron signals for a gluino LSP, or monotop signals from $\tilde{t}\rightarrow t\,\tilde{\chi}^{0}$ for a quasi stable neutralino LSP flying away. We will not consider these scenarios here~\cite{Monotops}.} 

\subsection{Squarks lighter than the gluino}

As shown in Fig.~\ref{DecaysFig}, the situation is rather involved in this case. As a starting point, let us imagine that all the squarks are precisely degenerate in mass while both the gluino and neutralino are heavier. There are then neither mixings nor transitions among the squarks. Instead, the right-handed squarks decay directly to quarks thanks to the RPV couplings, while the left-handed squarks need to go through a virtual gluino or neutralino to do so (see Fig.~\ref{Signals}$d$):
\begin{align}
\Gamma\Big(\tilde{u}_{R}^{I}  &
	\rightarrow\bar{d}^{J}\bar{d}^{K}
	\Big)
	\approx\frac{M_{\tilde{u}_{R}^{I}}}{8\pi}|\bm{\lambda}_{IJK}^{\prime\prime}|^{2}\;,\\
\Gamma\Big(\tilde{u}_{R}^{I}\,(\tilde{u}_{L}^{I})  & 
	\rightarrow u^{I}+\tilde{g}^*
	\rightarrow u^{I}t\,d\,s\,(u^{I}\bar{t}\,\bar{d}\,\bar{s})
	\Big)
	\approx\frac{\alpha_{S}^{2}M_{\tilde{u}_{L,R}^{I}}}{6600\pi^{3}}\times|\bm{\lambda}_{tds}^{\prime\prime}|^{2}\times\frac{M_{\tilde{q}}^{2}}{M_{\tilde{g}}^{2}}\;,
\label{uL1}%
\end{align}
and similarly for $\tilde{d}_{L,R}^{I}$, where again $m_{t}/M_{\tilde q}$ and higher powers of $M_{\tilde{q}}/M_{\tilde{g}}$ are neglected. Remark that even though the Majorana gluino decays to $t\,d\,s$ and $\bar{t}\,\bar{d}\,\bar{s}$ with equal probability, $\tilde{q}_{R}^{I}$ decays mostly to top quark and $\tilde{q}_{L}^{I}$ to anti-top quark because the $\tilde{q}_{R}^{I}\rightarrow q^{I}\bar{t}\,\bar{d}\,\bar{s}$ and $\tilde{q}_{L}^{I}\rightarrow q^{I}t\,d\,s$ rates scale like $M_{\tilde{q}}^{4}/M_{\tilde{g}}^{4}$ instead of $M_{\tilde{q}}^{2}/M_{\tilde{g}}^{2}$ (more details, as well as the rates for the neutralino-induced processes can be found in appendix~\ref{App2B} and~\ref{App4B}). Numerically, for $M_{\tilde{g}}\approx1$~TeV and $M_{\tilde{q}}\approx 300$~GeV, the four-body decay width is larger than about $(10^{-8}$ GeV$)\times|\bm{\lambda}_{tds}^{\prime\prime}|^{2}$, see Fig.~\ref{FigAppLT} in the appendix. So, the squarks are not viable R-hadron candidates when $\bm{\lambda}^{\prime\prime}$ follows the full MFV hierarchy. Note however that the two-body decay rates of the right-handed squarks span several orders of magnitude. In particular, for light flavors, the four-body channels sometimes dominate when $M_{\tilde{g}}$ is not too large. This is particularly true when the neutralino is lighter than squarks, in which case most of them decay first to neutralinos, which then decay to $t\,d\,s$ or $\bar{t}\,\bar{d}\,\bar{s}$.

The introduction of realistic squark mass splittings complicates this picture. Under MFV, the squark soft-breaking terms are fixed in terms of the Yukawa couplings as~\cite{MFV}
\begin{equation}
\begin{aligned}
\mathbf{m}_{Q}^{2}  &  = m_{0}^{2}\;
	\left[\mathbf{1}\oplus
	\mathbf{Y}_{d}^{\dagger}\mathbf{Y}_{d}\oplus
	\mathbf{Y}_{u}^{\dagger}\mathbf{Y}_{u}\oplus...\right]\;,\\
\mathbf{m}_{U}^{2}  &  = m_{0}^{2}
	\left(\mathbf{1}\oplus
	\mathbf{Y}_{u}
		\left[\mathbf{1}\oplus
		\mathbf{Y}_{d}^{\dagger}\mathbf{Y}_{d}\oplus
		\mathbf{Y}_{u}^{\dagger}\mathbf{Y}_{u}\oplus...\right]
	\mathbf{Y}_{u}^{\dagger}
	\right)\;,\\
\mathbf{m}_{D}^{2}  &  = m_{0}^{2}\;
	\left(\mathbf{1}\oplus
	\mathbf{Y}_{d}
		\left[\mathbf{1}\oplus
		\mathbf{Y}_{d}^{\dagger}\mathbf{Y}_{d}\oplus
		\mathbf{Y}_{u}^{\dagger}\mathbf{Y}_{u}\oplus...\right]
	\mathbf{Y}_{d}^{\dagger}
	\right)\;,\\
\mathbf{A}_{u}  &  = A_{0}\;\mathbf{Y}_{u}
	\left[\mathbf{1}\oplus
	\mathbf{Y}_{d}^{\dagger}\mathbf{Y}_{d}\oplus
	\mathbf{Y}_{u}^{\dagger}\mathbf{Y}_{u}\oplus...\right]\;,\\
\mathbf{A}_{d}  &  = A_{0}\;\mathbf{Y}_{d}
	\left[\mathbf{1}\oplus
	\mathbf{Y}_{d}^{\dagger}\mathbf{Y}_{d}\oplus
	\mathbf{Y}_{u}^{\dagger}\mathbf{Y}_{u}%
	\oplus...\right]\;.
\end{aligned}
\label{MFVsoft}%
\end{equation}
As in Eq.~(\ref{UDD2}), $\oplus$ indicates that arbitrary order one coefficients are understood for each term. In this way, flavor changing effects are consistently tuned by the Cabibbo-Kobayashi-Maskawa (CKM) matrix, and supersymmetric contributions to the flavor-changing neutral currents end up sufficiently suppressed to pass experimental bounds.

\begin{table}[t]
\renewcommand{\arraystretch}{1.2}
\begin{equation*}
\begin{array}[c]{ccccc}
\hline
& \multicolumn{2}{c}{\text{Full}}	& \multicolumn{2}{c}{\text{Holomorphic}} \\\hline
\tan\beta			& 5		& 50		& 5		& 50 \\\hline
\Gamma(\tilde{q})_{\min}^{4-body}	& 10^{-10}	& 10^{-8}	& 10^{-18}	& 10^{-14} \\\hline
\Gamma(\tilde{u}_{R})		& 10^{-9}	& 10^{-7}	& 10^{-15}	& 10^{-11} \\
\Gamma(\tilde{u}_{L})^{dir}	& 10^{-19}	& 10^{-17}	& 10^{-25}	& 10^{-21} \\
\Gamma(\tilde{u}_{L})^{mix}	& 10^{-10}	& 10^{-4}	& 10^{-18}	& 10^{-10} \\\hline
\Gamma(\tilde{c}_{R})		& 10^{-7}	& 10^{-5}	& 10^{-11}	& 10^{-7} \\
\Gamma(\tilde{c}_{L})^{dir}	& 10^{-12}	& 10^{-10}	& 10^{-16}	& 10^{-12} \\
\Gamma(\tilde{c}_{L})^{mix}	& 10^{-8}	& 10^{-2}	& 10^{-16}	& 10^{-8} \\\hline
\Gamma(\tilde{t}_{R})		& 0.1		& 10		& 10^{-9}	& 10^{-5} \\
\Gamma(\tilde{t}_{L})		& 0.1		& 10		& 10^{-9}	& 10^{-5} \\\hline
\end{array}%
\qquad
\begin{array}[c]{ccccc}
\hline
& \multicolumn{2}{c}{\text{Full}}	& \multicolumn{2}{c}{\text{Holomorphic}}\\\hline
\tan\beta			& 5		& 50		& 5		& 50 \\\hline
\Gamma(\tilde{q})_{\min}^{4-body}	& 10^{-10}	& 10^{-8}	& 10^{-18}	& 10^{-14}\\\hline
\Gamma(\tilde{d}_{R})		& 0.1		& 10		& 10^{-11}	& 10^{-7}\\
\Gamma(\tilde{d}_{L})^{dir}	& 10^{-10}	& 10^{-6}	& 10^{-20}	& 10^{-14}\\
\Gamma(\tilde{d}_{L})^{mix}	& 10^{-15}	& 10^{-11}	& 10^{-17}	& 10^{-11}\\\hline
\Gamma(\tilde{s}_{R})		& 0.1		& 10		& 10^{-9}	& 10^{-5}\\
\Gamma(\tilde{s}_{L})^{dir}	& 10^{-7}	& 10^{-3}	& 10^{-15}	& 10^{-9}\\
\Gamma(\tilde{s}_{L})^{mix}	& 10^{-12}	& 10^{-8}	& 10^{-14}	& 10^{-8}\\\hline
\Gamma(\tilde{b}_{R})		& 10^{-7}	& 10^{-5}	& 10^{-10}	& 10^{-5}\\
\Gamma(\tilde{b}_{L})		& 10^{-10}	& 10^{-6}	& 10^{-12}	& 10^{-6}\\\hline
\end{array}
\end{equation*}
\caption{Order of magnitude estimates of the squark decay widths (in GeV) when only the RPV modes are kinematically open, setting all squark masses at $300$~GeV, and assuming the MFV hierarchies shown in Table~\ref{Hierarchies}. The four-body decay widths quoted in the first line, corresponding to Eq.~(\ref{uL1}) with a gluino mass of $1$~TeV, are universal and represent the upper limits for all the squark lifetimes. The superscripts $dir$ refers to the direct $\tilde{q}^I_L \rightarrow \tilde{q}^I_R\rightarrow\bar{q}^J \bar{q}^K$ decay channel, Eq.~(\ref{uL2}), and $mix$ to those allowed by the flavor mixings in the squark soft-breaking terms once the MFV prescription is imposed, Eq.~(\ref{uL3}). For $\tilde{t}_L$ and $\tilde{b}_L$, these two mechanisms yield the same widths. Note that the $\tan\beta$ scaling of the partial widths can be easily inferred from the values given for $\tan\beta = 5$ and $50$.}
\label{TableLT}
\end{table}

The mass spectra induced by the MFV prescription are similar to those obtained starting with universal GUT boundary conditions but for two crucial differences~\cite{MFVRGE}. First, because of the $\mathcal{O}(1)$ coefficients, the leading flavor-blind terms of $\mathbf{m}_{Q}^{2}$, $\mathbf{m}_{U}^{2}$, and $\mathbf{m}_{D}^{2}$ need not be identical at any scale. Second, the third generation squark masses can be significantly split from the first two, especially when $\tan\beta$ is large. This originates from the hierarchy of $\mathbf{Y}_{u}^{\dagger}\mathbf{Y}_{u}$ and $\mathbf{Y}_{d}^{\dagger}\mathbf{Y}_{d}$: both have as largest entry their $33$ component. A typical MFV spectrum at moderate $\tan\beta$ is thus made of the quasi degenerate sets $\{\tilde{u}_{L},\tilde{c}_{L},\tilde{d}_{L},\tilde{s}_{L},\tilde{b}_{L}\}$, $\{\tilde{u}_{R},\tilde{c}_{R}\}$, $\{\tilde{d}_{R},\tilde{s}_{R},\tilde{b}_{R}\}$, together with the stop eigenstates $\tilde{t}_{1,2}$ which are split from their flavor partners by the large $\mathbf{A}_{u}^{33}$. When $\tan\beta$ is large, the sbottom mass eigenstates $\tilde{b}_{1,2}$ are also split from their flavor partners. Note that such a large stop mixing may actually be required to push the lightest Higgs boson mass up to about $125$~GeV~\cite{MaxMixetal}.

The MFV prescription for the squark mass terms impacts the decay chains in three ways. First, $\tilde{t}\rightarrow\tilde{b}\,W$ or $\tilde{b}\rightarrow\tilde{t}\,W$ may possibly open. Weak decays are irrelevant for the other squark flavors because $\tilde{u}_{L}$, $\tilde{c}_{L}$, $\tilde{d}_{L}$, and $\tilde{s}_{L}$ are essentially degenerate, and their $LR$ mixings are small. Note that when MFV is active, flavor-changing weak decays of the $\tilde{t}$ and $\tilde{b}$ are suppressed by the small CKM angles, and can be safely neglected. Second, squarks can cascade decay among themselves through the three-body $\tilde{q}\rightarrow q\,\bar q^{\prime}\tilde{q}^{\prime}$ processes mediated by a virtual\footnote{This remains true when the gluino or neutralino is real with a mass lying somewhere in-between the squark states.} gluino or neutralino. This is relevant only for those squarks having suppressed RPV decays like for example $\tilde{u}_{L,R}\rightarrow u\,\bar d\,\tilde{d}_{R}$ if\footnote{Here and in the following, we denote a specific squark mass hierarchy in terms of the corresponding soft mass term hierarchy, even though it is understood that squark masses do not depend only on these parameters.}  $(\mathbf{m}_{D}^{2})^{11}<(\mathbf{m}_{Q,U}^{2})^{11}$. Third, the RPV two-body decay modes open up for the left-handed squarks thanks to the non zero $(\mathbf{A}_{u,d})^{II}$, and to the flavor mixings present in $\mathbf{m}_{Q}^{2}$ and $\mathbf{A}_{u,d}$. Taking the up-type squarks for definiteness and assuming $\bm{\lambda}_{tds}^{\prime\prime}$ dominates, their partial decay widths are
\begin{align}
\Gamma\Big(\tilde{u}_{L}^{I}  &  \rightarrow\tilde{u}_{R}^{I}\rightarrow\bar{d}^{J}\bar{d}^{K}\Big)^{dir}	\approx\frac{M_{\tilde{u}_{L}^{I}}}{8\pi}\left|\frac{m_{u^{I}}}{v_{u}}\bm{\lambda}_{IJK}^{\prime\prime}\right| ^{2}\;,\;
\label{uL2}\\
\Gamma\Big(\tilde{u}_{L}^{I}  &  \rightarrow\tilde{t}_{R}\rightarrow\bar{d}\,\bar{s}\Big)^{mix}
	\approx\Gamma\Big(\tilde{u}_{L}^{I}
		\rightarrow\tilde{t}_{L}
		\rightarrow\tilde{t}_{R} 
		\rightarrow\bar{d}\,\bar{s}\Big)^{mix}
	\approx \frac{M_{\tilde{u}_{L}^{I}}}{8\pi}
		\left|\frac{m_{t}}{v_{u}} \frac{m_{b}^{2}}{v_{d}^{2}} V_{Ib} V_{tb}^{\ast} \bm{\lambda}_{tds}^{\prime\prime} \right|^{2}\;,
\label{uL3}
\end{align}
where we set $m_{0}\approx A_{0}$. In the $I=1$ case, the direct channel is extremely suppressed by the tiny left-right mixing $\mathbf{A}_{u}^{11}\sim m_{u}/v_{u}$ and RPV couplings $\bm{\lambda}_{uJK}^{\prime\prime}$. By contrast, the indirect channel tuned by $\bm{\lambda}_{tds}^{\prime\prime}$ becomes available at the relatively modest cost of $|V_{ub}V_{tb}^{\ast}|\approx10^{-3}$ thanks to the flavor mixings in $\mathbf{m}_{Q}^{2}$ and $\mathbf{A}_{u}$ (specifically, to the $\mathbf{Y}_{d}^{\dagger}\mathbf{Y}_{d}$ terms in Eq.~(\ref{MFVsoft})). Note that $\mathbf{Y}_{d}^{\dagger}\mathbf{Y}_{d}$ is proportional to $m_{b}^{2}/v_{d}^{2}$, so $\Gamma(\tilde{u}_{L})^{mix}$ has a very strong $\tan^6\beta$ dependence once accounting for the $\tan\beta$ scaling of $\bm{\lambda}^{\prime\prime}$ (this further increases to a $\tan^8\beta$ dependence in the holomorphic case). It actually ends up larger than $\Gamma(\tilde{u}_{R})$ when $\tan\beta\gtrsim10$ (see Table~\ref{TableLT}). Indeed, a similar decay mechanism for $\tilde{u}_{R}$ is never competitive once MFV is imposed because $\mathbf{Y}_{d}^{\dagger}\mathbf{Y}_{d}$ occurs in $\mathbf{m}_{U}^{2}$ only sandwiched between $\mathbf{Y}_{u}$ and $\mathbf{Y}_{u}^{\dagger}$. So, $(\mathbf{m}_{U}^{2})^{13}$ is proportional to the tiny up-quark mass and $\tilde{u}_{R}\rightarrow\tilde{t}_{R}\rightarrow\bar{d}\,\bar{s}$ is very suppressed.

As said above, MFV is compatible with a stop LSP, since it naturally allows for a large splitting of the third generation squarks. In that case, most decay chains still end with a top quark, see Fig.~\ref{DecaysFig}. Indeed, though the RPV decay $\tilde{t}\rightarrow jj$ is top-less and very fast, the stops arise mostly from the flavor-conserving decays of heavier sparticles, and are thus produced together with top quarks. For example, the gaugino decays exclusively to $t,\bar{t}+jj$ independently of whether it is a true LSP or a yet lighter stop is present.

\subsection{Combining sparticle production mechanisms with decay chains}

With the full MFV hierarchy, most decay chains end up with a top quark (see Fig.~\ref{DecaysFig}). Further, without large mass splittings, the sparticle decay widths are large enough to avoid R-hadron constraints. Actually, most decays are even way too fast to leave displaced vertices (see Fig.~\ref{FigLongLife}).\footnote{Note, though, that a colored LSP would live long enough to hadronize.} So, given the production mechanisms of Eq.~(\ref{LHCProd}), the supersymmetric processes can be organized into two broad classes. If the first-generation squarks are heavier than the gluino, then there are no final states made entirely of light-quark jets:
\begin{equation}
M_{\tilde{g}}<\mathbf{m}_{Q,U,D}^{2}:\;\;g\,g\rightarrow \tilde{g}\,\tilde{g}\rightarrow (t\,t,\bar{t}\,\bar{t}\,)+4j/6j/8j\;\;,
\end{equation}%
with the number of jets increasing when gluinos first cascade decay to neutralinos. Note that we already discarded the $t\;\bar{t}+\text{jets}$ final state, since it would correspond to a $\Delta B = 0$ process.

Conversely, if the squarks are lighter than the gluinos, then most but not all decay chains terminate with a top quark. So, most of the processes initiated by the proton $u$ and/or $d$ quarks lead to same-sign top-quark pairs:%
\begin{equation}
\begin{array}{ll}
\mathbf{m}_{D}^{2} < \mathbf{m}_{Q,U}^{2},M_{\tilde{g}}: 
  &  d\,d\rightarrow \tilde{d}_{R}\,\tilde{d}_{R}\rightarrow \bar{t}\,\bar{t}+2j\;, \\[2mm]
\mathbf{m}_{Q}^{2}<\mathbf{m}_{D,U}^{2},M_{\tilde{g}}:
  & d\,d\rightarrow \tilde{d}_{L}\,\tilde{d}_{L}\rightarrow \bar{t}\,\bar{t}+2j/4j/6j\;, \\ 
  & u\,d\rightarrow \tilde{u}_{L}\,\tilde{d}_{L}\rightarrow \bar{t}\,\bar{t}+4j/6j\;;\;\;t+3j\;,\\ 
  & u\,u\rightarrow \tilde{u}_{L}\,\tilde{u}_{L}\rightarrow \bar{t}\,\bar{t}+6j\;;\;\;4j\;, \\[2mm]
\mathbf{m}_{U}^{2}<\mathbf{m}_{Q,D}^{2},M_{\tilde{g}}: 
  & u\,u\rightarrow \tilde{u}_{R}\,\tilde{u}_{R}\rightarrow t\,t+6j\;;\;\;4j\;,%
\end{array}
\label{SquTT}
\end{equation}
where we neglected the suppressed decay $\tilde{u}_{L}\,\tilde{u}_{L}\rightarrow t\,t+6j$ and $\tilde{u}_{R}\,\tilde{u}_{R}\rightarrow \bar{t}\,\bar{t}+6j$ (see the discussion in appendix~\ref{App4B}). Again, the number of jets increases when at least one neutralino is lighter than the squarks. In these equations, the comparisons between $\mathbf{m}_{Q}^{2}$, $\mathbf{m}_{U}^{2}$, and $\mathbf{m}_{D}^{2}$ are understood to apply to their 11 and 22 entries which give, to an excellent approximation, the first two generation squark masses (see Eq.~(\ref{MFVsoft})).

Whatever sparticle production mechanism dominates, the precise production rate of same-sign top-quark pairs depends on whether the squarks, when they are not the lightest, prefer to undergo their RPV decay or, instead, to cascade decay down to other squarks, which in turn may or may not produce same-sign top pairs. For example, when $\mathbf{m}_{D}^{2} < \mathbf{m}_{Q,U}^{2}$, it is quite possible that $\tilde{u}_L$, $\tilde{u}_R$, and $\tilde{d}_L$ all decay into $\tilde{d}_R$, which then decays to $\bar{t}+j$. Conversely, when $\mathbf{m}_{U}^{2}<\mathbf{m}_{Q,D}^{2}$ and $M_{\tilde{g},\tilde{\chi}^{0}}$ is large, we may be in a situation where all of them but $\tilde{d}_R$ cascade down to $\tilde{u}_R$, which then produces two jets. In this case, only $\tilde{d}_R\tilde{d}_R$ produces top pairs. So, depending on the MSSM mass spectrum, the amount of same-sign top pairs can span more than an order of magnitude.

With the holomorphic MFV hierarchy, the above picture remains valid, at least qualitatively. The decay chains still mostly end up with top quarks and the amount of same-sign top pairs emerging from the production mechanisms of Eq.~(\ref{LHCProd}) is not much affected. There are four differences worth noting though. First, some light-quark jets are replaced by $b$ jets in all final states. Second, the branching ratios for the three left-squark decay modes, Eq.~(\ref{uL1}),~(\ref{uL2}), and~(\ref{uL3}), are affected, hence the decay chains do not necessarily follow the same paths as with the full MFV hierarchy. Third, all $\bm{\lambda}^{\prime\prime}$ couplings are now much smaller than $\alpha$, so the direct RPV decays are systematically subdominant whenever $\tilde{q}\rightarrow q\,\tilde{\chi}^0_1$ or $\tilde{g}\rightarrow q\,\bar{q}\,\tilde{\chi}^0_1$ are kinematically open (assuming $\tilde{\chi}^0_1$ is essentially a bino). Same-sign top-quark pairs still arise, but are in general accompanied by many more jets. Finally, a light LSP, whether it is a gluino, neutralino, or a squark, can have a large lifetime when $\tan\beta$ is small, even for moderate mass hierarchies (see Fig.~\ref{FigLongLife}). This is the only corner of parameter space in which R-hadron constraints could play a role.

Specifically, looking at Table~\ref{TableLT}, the lifetimes are always below about 1~$\mu$s. This is rather short, so we should use the bounds the Atlas collaboration sets using the inner detector only~\cite{StablesQ}, which requires the total width of the sparticle to be below about $10^{-16}$~GeV (see Fig.~\ref{FigLongLife}). Such a lifetime for the squarks is a priori possible only for the $\tilde{u}_L$ and $\tilde{d}_L$. It further requires $\tan\beta\lesssim 10$ and $A_0\lesssim m_0$, otherwise the two-body decay rates Eq.~(\ref{uL3}) are above $10^{-16}$~GeV even for $M_{\tilde{u}_L,\tilde{d}_L}$ as low as $300$~GeV. Both these conditions appear contradictory to the requirements of a rather large Higgs boson mass~\cite{MaxMixetal}, which asks for a not too small $\tan\beta$ and relatively large trilinear terms. So, even with the holomorphic MFV hierarchy, squarks do not appear viable as R-hadron candidates. Turning to the gluino, although its lifetime can always be made long enough by increasing the squark masses, this nevertheless requires pushing them to very large values. For $\tan\beta=5$ and $M_{\tilde{g}}=(250,500,1000)$~GeV, the gluino width is below $10^{-16}$~GeV for $M_{\tilde{q}}\gtrsim (1,5,13)$~TeV. This is the range excluded by the Atlas bound. Note that the squark and gluino lifetimes increase if their mass is below $m_t$, since this shuts down the dominant RPV decay mode. But the Atlas bounds on the squark and gluino masses are already well above $m_t$, so this region is excluded. We thus conclude that the R-hadron constraints play no role over the mass range over which the dilepton signal will be probed in the following, which goes from $M_{\tilde{g}},M_{\tilde{q}}\approx 200$ to about $1100$~GeV.

%%%%%%%%%%%%%%%%%%%%%%%%%%%%%%%%%%%%%%%%%%%%%%%%%%%%%%%%%%%%%%%%%%%%
\section{Simplified mass spectrum and analysis strategy\label{EffDecCh}}

In view of the complexity of the decay chains discussed in the previous section, it is very desirable to design a simplified analysis strategy. For instance, the exact squark decay chains depend on the many MSSM parameters tuning the squark masses and the three decay modes of Eq.~(\ref{uL1}),~(\ref{uL2}), and~(\ref{uL3}), so one should in principle perform a full scan over these parameters.

The situation is, however, more simple than it seems. Indeed, given that there are only two broad classes of decay chains, it is possible to simulate them generically by introducing only two mass scales, $M_{\tilde{g}}$ and $M_{\tilde{q}}$, with $M_{\tilde{q}}$ denoting the first generation squark mass scale. Though not immediately apparent, this is sufficient to encompass in a very realistic fashion the dominant decay chains for most mass spectra. Indeed:
\begin{description}
\item[\fbox{$M_{\tilde{g}}<M_{\tilde{q}}$}] This sector describes generically the situation where squarks are heavier than the gluino, and is dominated by the $g\,g\rightarrow\tilde{g}\,\tilde{g}$ production mechanism. Assuming neutralinos are heavier, each gluino then decays exclusively to $(t,\bar{t}\,)+2j$. There are as many $t\,t$ as $\bar{t}\,\bar{t}$ pairs so the lepton charge asymmetry vanishes,
\begin{equation}
\sigma(p\,p\rightarrow\tilde{g}\,\tilde{g} \rightarrow \bar{t}\,\bar{t}+4j):
\sigma(p\,p\rightarrow\tilde{g}\,\tilde{g} \rightarrow t\,t+4j)
\approx1:1\;.
\label{proc1}
\end{equation}
Note that $\sigma(g\,g \rightarrow \tilde{g}\,\tilde{g}) = 2\times\sigma(g\,g \rightarrow \tilde{g}\,\tilde{g} \rightarrow (t\,t , \bar{t}\,\bar{t})+4j)$, expected from the Majorana nature of the gluinos, is not always strictly true, especially when the gluino width is large~\cite{Berdine07}. The reason is to be found in the chirality of the RPV and gluino couplings, which selects either the $\slashed p$ or the $M_{\tilde{g}}$ terms of the gluino propagators (see the discussion in appendix~\ref{App4B}). The signal is similar if the neutralino replaces the gluino as LSP, with the same-sign top quarks produced through $g\,g \rightarrow \tilde{g}\,\tilde{g} \rightarrow \tilde{\chi}^{0}\tilde{\chi}^{0}+4j\rightarrow(t\,t,\bar{t}\,\bar{t}\,)+8j$. The top-quark energy spectra would then be slightly softer because of the longer decay chains. Our bounds on the gluino mass are, in this case, only approximate. On the other hand, the precise squark mass spectrum is almost completely irrelevant since it affects only the gluino (or neutralino) lifetime, not its decay modes. This remains true even if the stop is the LSP. When the other squarks are heavier than the gauginos, for instance, the gluinos almost exclusively decay through $\tilde{g} \rightarrow t\,\bar{\tilde{t}},\bar{t}\,\tilde{t} \rightarrow(t,\bar{t}\,)+2j$.

\item[\fbox{$M_{\tilde{g}}>M_{\tilde{q}}$}] This sector describes generically the situation where first-generation squarks are lighter than gauginos. Looking at all the processes in Eq.~(\ref{SquTT}), the crucial observation is that $d\,d \rightarrow \tilde{d}_{R}\,\tilde{d}_{R} \rightarrow \bar{t}\,\bar{t}+2j$ is always active, while the other squark intermediate states may or may not lead to same-sign top pairs, depending on the MSSM parameters. So, to account for a large range of possibilities, we span from the pessimistic situation where $p\,p \rightarrow \tilde{d}_{R}\,\tilde{d}_{R} \rightarrow \bar{t}\,\bar{t}+2j$ is the only top-pair producing channel, to the optimistic situation where $d\,d\rightarrow\tilde{d}_{L}\tilde{d}_{L}\rightarrow\bar{t}\,\bar{t}+2j$ and $d\,d\rightarrow\tilde{d}_{R}\tilde{d}_{L}\rightarrow\bar{t}\,\bar{t}+2j$ are also active, with $\tilde{d}_{R,L}$ both of mass $M_{\tilde{q}}$ and with unit branching fraction to $\bar t+j$. The much longer $\tilde{d}_{L}$ lifetime is not directly relevant, at least as long as it decays within the detector.\footnote{With the holomorphic hierarchy, when $\tan\beta \lesssim 15$ (or a bit lower if $A_0 > m_0$ at the TeV scale), the $\tilde{d}_{L}$ lifetime could be above about $0.1$~ns (see Fig.~\ref{FigLongLife}). At that point, the identification of top-quark pairs through same-sign leptons plus $b$-jets starts loosing efficiency because the $b$ tagging requires a secondary vertex no farther than a few centimeters away from the primary one~\cite{Buchkremer:2012dn}.} Note that the $p\,p \rightarrow \tilde{d}_{R}\,\tilde{d}_{L}$ channel would be the only one to survive if gluinos were Dirac particles~\cite{DiracGlue}. In any case, since the $\bar{d}$ proton PDF is significantly smaller than that of the $d$, the lepton charge asymmetry is close to maximally negative:
\begin{equation}
\sigma(p\,p \rightarrow \tilde{d}_{R,L}\,\tilde{d}_{R,L} \rightarrow \bar{t}\,\bar{t}+2j):
\sigma(p\,p \rightarrow \bar{\tilde{d}}_{R,L}\,\bar{\tilde{d}}_{R,L} \rightarrow t\,t+2j)
\approx1:0\;.
\label{proc2}
\end{equation}
In principle, the number of top pairs could further be increased by nearly an order of magnitude if up quarks come into play. For simplicity and since these modes give rise to softer final states of higher jet multiplicity, we prefer to disregard them. In addition, realistic situations probably lie somewhere between our pessimistic and optimistic settings, with some top pairs coming from both $\tilde{d}_{L}$ and $\tilde{u}_{L,R}$ but with $\mathcal{B}(\tilde{d}_{L}\rightarrow\bar{t}+j)$ and $\mathcal{B}(\tilde{u}_{L,R}\rightarrow\bar{t}+3j)<1$. Note also that, if the contribution of $u\,u \rightarrow \tilde{u}\,\tilde{u} \rightarrow t\,t+6j$ is significant (for intermediate $\tilde{u}_{L}$, this requires a rather light gaugino), or if all the four-body final states are strongly favored by a light neutralino, the lepton charge asymmetry could be somewhat diluted.

\item[\fbox{$M_{\tilde{g}}\approx M_{\tilde{q}}$}] In this region, in addition to $g\,g \rightarrow \tilde{g}\,\tilde{g}$ and $d\,d \rightarrow \tilde{d}_{R}\,\tilde{d}_{R}$, the mixed production $g\,d \rightarrow \tilde{g}\,\tilde{d}_{R} \rightarrow \bar{t}\,\bar{t}+3j$ is competitive. In the optimistic case, an equal amount of top-quark pairs is produced through the $g\,d \rightarrow \tilde{g}\,\tilde{d}_{L} \rightarrow \bar{t}\,\bar{t}+3j$ process. As for the $d\,d \rightarrow \tilde{d}\,\tilde{d}$ processes, the proton PDF strongly favors negative lepton pair productions:
\begin{equation}
\sigma(p\,p \rightarrow \tilde{g}\,\tilde{d}_{R,L} \rightarrow\bar{t}\,\bar{t}+3j):
\sigma(p\,p \rightarrow \tilde{g}\,\bar{\tilde{d}}_{R,L} \rightarrow t\,t+3j)
\approx1:0\;.
\label{proc3}
\end{equation}
Compared to the other cases, it should be stressed that the decay chains in the $M_{\tilde{g}} \approx M_{\tilde{q}}$ region can be rather complicated. Indeed, squarks are not precisely degenerate in mass, so this region includes compressed spectra with the gluino (or neutralino) mass lying in-between squark masses. Overall, the amount of top pairs should not be very significantly reduced, but their production may proceed through rather indirect routes. For instance, one of the worst case scenario occurs when $\mathbf{m}_{U}^{2} \lesssim M_{\tilde{g}} \lesssim \mathbf{m}_{D,Q}^{2}$. The $\tilde{d}_{R,L} \rightarrow d\,\tilde{g}$ decay competes with $\tilde{d}_{R,L} \rightarrow \bar{t}+j$ and $\tilde{g} \rightarrow u\,\tilde{u}, c\,\tilde{c}$ competes with $\tilde{g} \rightarrow t\,\tilde{t}$, thereby strongly depleting the amount of directly produced top pairs. At the same time, $u\,u \rightarrow \tilde{u}_{R}\,\tilde{u}_{R}$ more than replenishes the stock of top pairs since the four-body decay modes entirely dominate when $M_{\tilde{g}} \approx M_{\tilde{q}}$ (and there are more $u$ quarks than $d$ quarks in the protons). This example shows that fixing the fine details of the mass spectrum is in principle compulsory to deal with compressed spectra, but also that our pessimistic estimates based only on the $g\,g \rightarrow \tilde{g}\,\tilde{g} \rightarrow \bar{t}\,\bar{t}+4j$, $g\,d \rightarrow \tilde{g}\,\tilde{d}_{R} \rightarrow \bar{t}\,\bar{t}+3j$, and $d\,d \rightarrow \tilde{d}_{R}\,\tilde{d}_{R} \rightarrow \bar{t}\,\bar{t}+2j$ production mechanisms should conservatively illustrate the experimental reach.
\end{description}

\begin{figure}[t]
\centering
\includegraphics[width=0.96\textwidth]{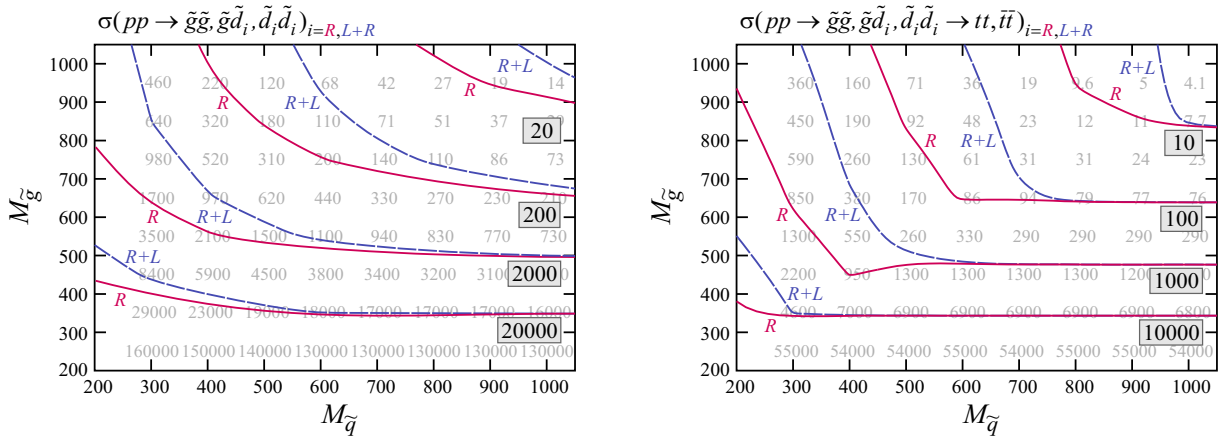}  
\caption{Sdown plus gluino production rate [fb] at the $8$~TeV LHC computed at leading order with MadGraph5, and the corresponding rate for the same-sign top-quark pair production, with or without active $\tilde{d}_L$ (the grid of numbers corresponds to the former case). When the top-quark pair arises from down-type squarks, the rate is not reduced since $\mathcal{B}(\tilde{d}_{R,L}\rightarrow \bar{t}\,\bar{s})=1$. When it arises from $g\,g\rightarrow \tilde{g}\,\tilde{g}$, the reduction is close to two since $\mathcal{B}(\tilde{g}\rightarrow t\,d\,s)=\mathcal{B}(\tilde{g}\rightarrow \bar{t}\,\bar{d}\,\bar{s})=1/2$.}%
\label{FigBare}%
\end{figure}

Thanks to the above simplifications, we only need to simulate the processes of Eqs.~(\ref{proc1}),~(\ref{proc2}), and~(\ref{proc3}). In practice, we use the \textsc{FeynRules}--\textsc{MadGraph5} software chain~\cite{Christensen:2008py, Alwall:2011uj} to produce leading order and parton level samples for the $8$ and $14$~TeV LHC. The squark and gluino masses $M_{\tilde{q}}$ and $M_{\tilde{g}}$ are then varied in the $200-1100$~GeV range while the neutralino, charginos, and sleptons are decoupled. In our analysis, we are not including the single-stop production mechanism (see Fig.~\ref{Production}). The reason is that it leads to same-sign top pairs only for a lighter gluino, in which case it is subleading compared to $g\,g\rightarrow \tilde{g}\,\tilde{g}$. We also neglect the subleading $q\,\bar{q}\rightarrow\tilde{g}\,\tilde{g}$ production mechanisms. If only the neutralino is lighter than the stop, there could be some same-sign top events only when $\tilde{t}\rightarrow jj$ is suppressed, like in the holomorphic case. We do not study that alternative here. We are also disregarding electroweak neutralino pair productions, or neutralino-induced squark production mechanisms, e.g., $d\,d\rightarrow \tilde{d}\,\tilde{d}$ via a neutralino (see Fig.~\ref{Production}). Both can generate same-sign top pairs, but are entirely negligible compared to the strong processes given the gluino mass range we consider here. So, neither the stop nor the neutralino are affecting the production mechanisms. In addition, we explained before that they do not affect the decay chains sufficiently to alter the same-sign top-quark pair production rate. So, for the time being, our signal is totally insensitive to both the stop and neutralino masses.

Throughout the numerical study, the RPV couplings are kept fixed to either $\bm{\lambda}_{tds}^{\prime\prime} = 0.1$ for the full MFV case, or $\bm{\lambda}_{tbs}^{\prime\prime}=10^{-3}$ and $\bm{\lambda}_{tds,tdb}^{\prime\prime}=10^{-4}$ for the holomorphic case, with all the smaller couplings set to zero. It should be clear though that the overall magnitude of these couplings does not play an important role. It affects the light sparticle lifetimes but not directly their branching ratios or their production rates. This is confirmed by the similarity of the results obtained in the next section with either the full or holomorphic hierarchy. Besides, since the sparticle widths play only a subleading role, we compute them taking all the squarks degenerate in mass.

This benchmark strategy is naturally suited to a two-dimensional representation in the $M_{\tilde{q}}-M_{\tilde{g}}$ plane (see Fig.~\ref{FigBare}). But, it must be stressed that even if this representation is seemingly similar to that often used for the search of the R-parity conserving MSSM, the underlying assumptions are intrinsically different and far less demanding in our case. Indeed, by using these two mass parameters and only a limited number of super-QCD production processes, our purpose is to estimate realistically the amount of same-sign top-quark pairs which can be produced. Crucially, no scenario with relatively light squarks and/or gluino could entirely evade producing such final states, and the experimental signals discussed in the next section are largely insensitive on how the top quarks are produced.

Finally, it should be mentioned that colored sparticle pair production is significantly underestimated when computed at leading order accuracy (compare Fig.~\ref{FigBare} with e.g. Ref.~\cite{ggNLO}), so the strength of our signal is certainly conservatively estimated. Our choice of working at leading order is essentially a matter of simplicity. Indeed, the whole processes are easily integrated within \textsc{MadGraph5}, including finite-width effects. In addition, our main goal here is to test the viability of our simplified theoretical framework and its observability at the LHC, so what really matters is the reduction in rate starting from Fig.~\ref{FigBare} and going through the experimental selection criteria. Of course, in the future, NLO effects should be included to derive sparticle mass bounds. But, given the pace at which experimental results in the dilepton channels are coming in, we refrain from doing this at this stage.

%%%%%%%%%%%%%%%%%%%%%%%%%%%%%%%%%%%%%%%%%%%%%%%%%%%%%%%%%%%%%%%%%%%%
\section{Same-sign dileptons at the LHC}

Both CMS~\cite{CMS7,CMS8} and ATLAS~\cite{Atlas7,Atlas8} have studied the same-sign dilepton signature at 7 and 8 TeV, and used it to set generic constraints on new physics contributions. Signal regions characterized by moderate missing energy, relatively high hadronic activity or jet multiplicity and one or two $b$ tags are expected to be the most sensitive to same-sign tops plus jets.

\subsection{Experimental backgrounds}

In these searches, irreducible and instrumental backgrounds have comparable magnitudes. Irreducible backgrounds with isolated same-sign leptons and $b$ jets arise from $t\bar{t}Z$ and $t\bar{t}\,W$ production processes. Their NLO cross sections~\cite{SMttZ,SMttW} amount respectively to $208$ and $232$~fb at the $8$~TeV LHC. The di- and tribosons ($W^{\pm}W^{\pm}$, $WZ$, $ZZ$; $WWW$, $WWZ$, $ZZZ$) plus jets productions also contribute, generally without hard $b$ jet and sometimes with a third opposite-sign lepton coming from a $Z$ boson. Positively charged dileptons dominate over negatively charged ones at the LHC when the net number of $W$ bosons (the number of $W^{+}$ minus the number of $W^{-}$) is non-vanishing. This feature is generic in the SM which communicates the proton-proton initial-state charge asymmetry to the final state.

Instrumental backgrounds arise from the misreconstruction (mainly in $t\bar{t}$ events) of
\begin{itemize}\setlength{\itemsep}{0mm}
\item (heavy) mesons decaying leptonically within jets,
\item hadrons as leptons,
\item asymmetric conversions of photons,
\item electron charges (if a hard bremsstrahlung radiation converts to a $e^{+}e^{-}$ pair in which the electron with a charge opposite to the initial one dominates).
\end{itemize}
The first three sources are often collectively referred to as \emph{fake leptons}. The important contribution of $b$ quark semi-leptonic decays in $t\bar{t}$ events with one top decaying semi-leptonically and the other hadronically is significantly reduced when (one or) several $b$ tags are required~\cite{CMS7}.

\begin{figure}[t]
\centering
\includegraphics[width=0.96\textwidth]{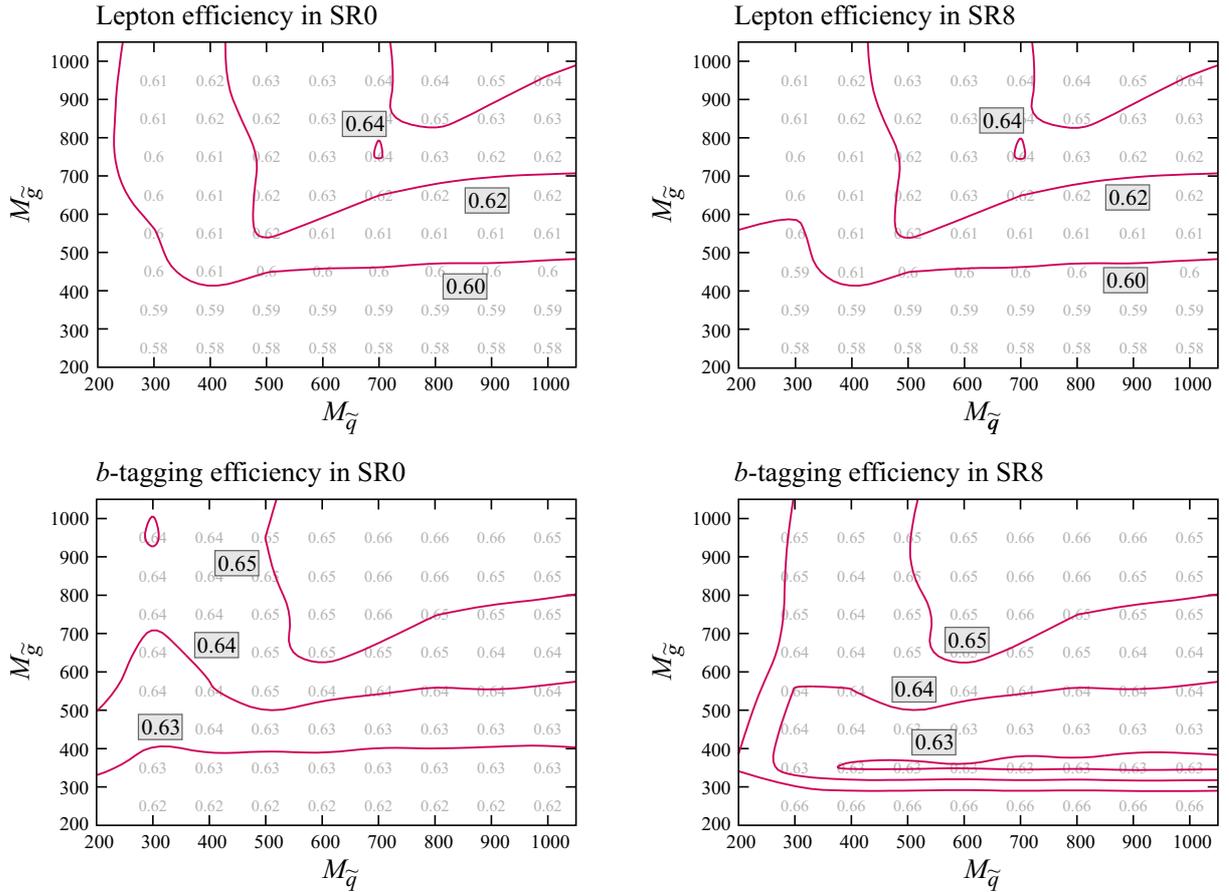}
\caption{Efficiencies for isolated lepton identification (top) and $b$ tagging (bottom) in signal region SR8 (left) and SR0 (right), using the $p_{T}$-dependent parametrization provided by CMS~\cite{CMS7,CMS8}. Since the RPV signal circumvents the significant drop in efficiencies for low $p_{T}$, these can be taken as constants in a good approximation. In our simulation, both of them are frozen at $60\%$.}%
\label{FigEff}%
\end{figure}

\subsection{Selection criteria}
We place ourselves in experimental conditions close to those of CMS, whose collaboration provides information (including efficiencies) and guidelines for constraining any model in an approximate way~\cite{CMS7,CMS8}. We ask for semi-leptonic decays of the top quarks to electrons or muons, and further require:
\begin{itemize}\setlength{\itemsep}{0mm}
\item two same-sign leptons with $p_{T}>20$~GeV and $|\eta|<2.4$,
\item at least two or four jets (depending on the signal region) with $p_{T}>40$~GeV and $|\eta|<2.4$,
\item at least two of these jets (three in one of the signal regions) to be $b$-tagged.
\end{itemize}
Still following CMS analyses, we define in Table~\ref{tab:sr} several signal regions (SR) with different cuts on the missing transverse energy $\slashed E_{T}$ and the transverse hadronic activity $H_{T}$. The selection of an isolated lepton is taken to have an efficiency of $60\%$ and the tagging of a parton level $b$ quark as a $b$ jet is fixed to be $60\%$ efficient too. These values have been chosen in view of the efficiencies obtained (see Fig.~\ref{FigEff}) using the $p_{T}$-dependent parametrization provided by CMS. Note that, for $b$ tagging, the value chosen is a few percent lower than those estimated in this way. With backgrounds under control, a higher number of isolated leptons from signal events could be selected by lowering the cut on their $p_{T}$ or by modifying the isolation requirement~\cite{Allanach:2012vj}.

To assess the goodness of our parton level approximate selection, we compared it (relaxing the same-sign condition for leptons) to the total acceptance in SR1 quoted by CMS for SM $t\bar{t}$ events with semi-leptonic top decays. Our total acceptance of $0.20\%$ (including top branching fractions) is compatible but lower than the $(0.29\pm 0.04)\%$ quoted by CMS~\cite{CMS8}. So, at this step, the strength of our signal is probably conservatively estimated.%

\newcommand{\myspace}{\hspace{1cm}\extracolsep{\fill}}
\begin{table}[t] \centering
\renewcommand{\arraystretch}{1.1}
\begin{tabular*}{\textwidth}{@{}r@{\qquad\extracolsep{\fill}}cc@{\myspace}cc@{\myspace}ccc@{\myspace}c}
\hline
	& SR0	& SR1	& SR4	& SR3	& SR8	& SR5	& SR6 	& SR7\\\hline
Min. number of $b$ tags
	& 2	& 2	& 2	& 2	& 2	& 2	& 2 	& 3\\
Min. number of extra jets
	& 0	& 0	& 2	& 2	& 2	& 2	& 2 	& 0\\
Cut on $H_T$ [GeV]
	& 80	& 80	& 200	& 200	& 320	& 320	& 320	& 200\\
Cut on $\slashed{E}_T$ [GeV]
	& 0	& 30	& 50	& 120	& 0	& 50	& 120 	& 50\\\hline
Limit on BSM events
	&  30.4	& 29.6	& 12.0	& 3.8	& 10.5	& 9.6	& 3.9	& 4.0\\\hline
\end{tabular*}%
\caption{Definitions of the signal regions used by CMS \cite{CMS8} for same-sign dilepton searches. For each of them, the $95\%$ CL upper limit on  beyond the SM (BSM) events is derived from $10.5$~fb$^{-1}$ of $8$~TeV data, assuming a $30\%$ uncertainty on signal efficiency and using the $CL_s$ method.}
\label{tab:sr}
\end{table}

\begin{figure}[t]
\centering
\includegraphics[width=0.96\textwidth]{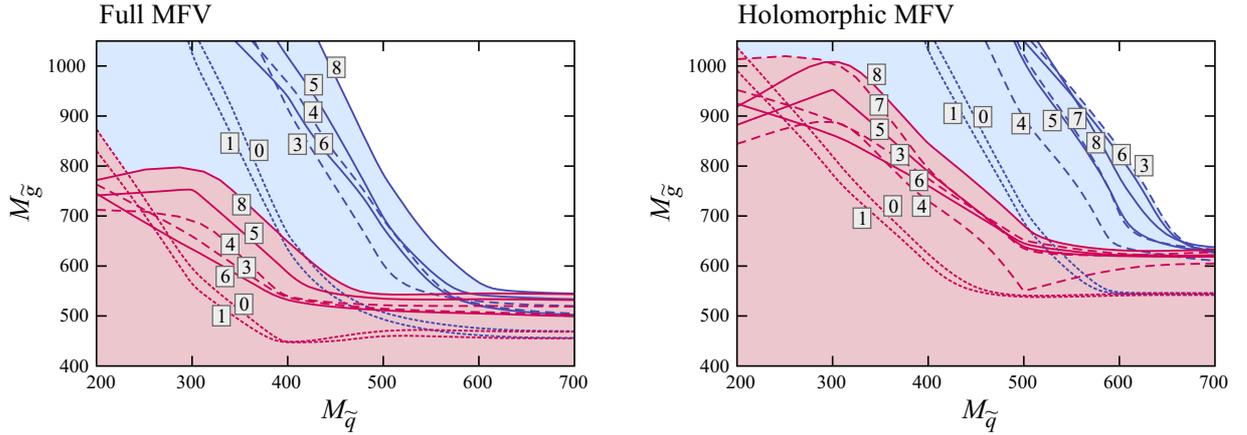}
\caption{$95\%$ CL exclusion contours in the $M_{\tilde{q}}-M_{\tilde{g}}$ plane derived from the CMS same-sign dilepton search~\cite{CMS8}. The lower red contours are obtained with the contribution of the $\tilde{d}_{R}$ only while the upper blue contours assume an equal contribution of $\tilde{d}_{R}$ and $\tilde{d}_{L}$ (i.e., with identical branching ratios to $\bar{t}\,\bar{s}$). Importantly, $M_{\tilde{q}}$ denotes the mass scale of the first generation squarks, $\tilde{u}_{R,L}$ and $\tilde{d}_{R,L}$. The presence of a light stop or a light neutralino does not significantly impact these exclusion regions. Finally, note that the R-hadron constraints in the holomorphic case are completely off the scale, requiring $M_{\tilde{q}}$ greater than at least a few TeV.}%
\label{FigLimits}%
\end{figure}

\subsection{Current constraints and prospects}

For several choices of squark and gluino masses, we count the number of events in each signal region and compare it with the $95\%$ CL limits set by CMS assuming a conservative $30\%$ uncertainty on the signal selection efficiency and using $10.5$~fb$^{-1}$ of $8$~TeV data~\cite{CMS8}. The corresponding exclusion contours in the $M_{\tilde{q}}-M_{\tilde{g}}$ plane are displayed in Fig.~\ref{FigLimits}.

\begin{figure}[t]
\centering
\includegraphics[width=0.96\textwidth]{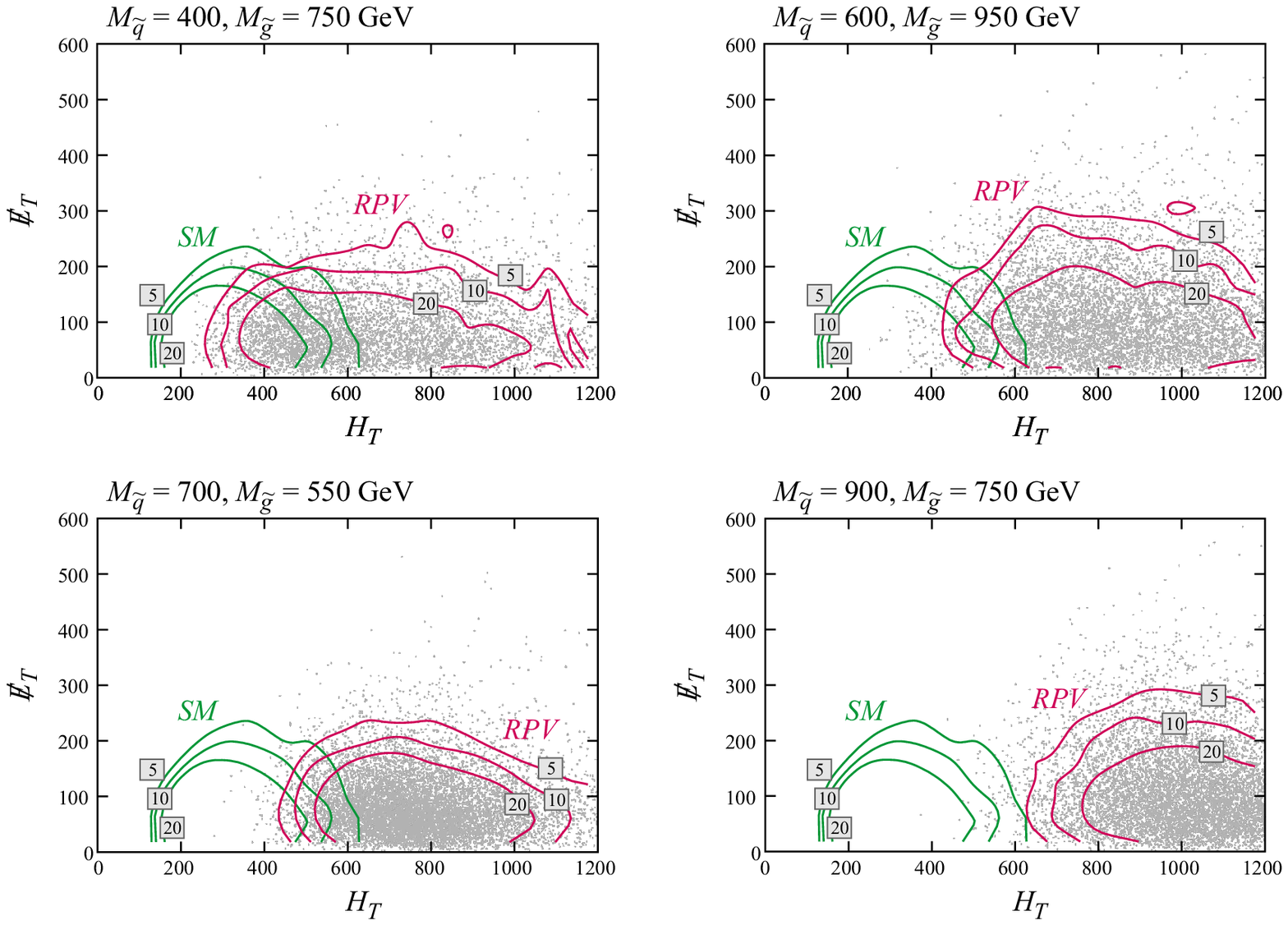}
\caption{Shape $1/\sigma\times d^{2}\sigma/dH_{T}d\slashed E_{T}\,$ $[100~GeV]^{-2}$ of the RPV signal in SR0 and in the full MFV hierarchy case. The $\tilde{d}_{L}$ contributions to the top pair production are not included here. For comparison, the shape of the SM $t\bar{t}\,W+t\bar{t}Z$ background is also shown. Those events are generated at leading order and parton level using \textsc{MadGraph5}~\cite{Alwall:2011uj}.}%
\label{HTET}%
\end{figure}

\begin{figure}[p]
\centering
\includegraphics[width=0.96\textwidth]{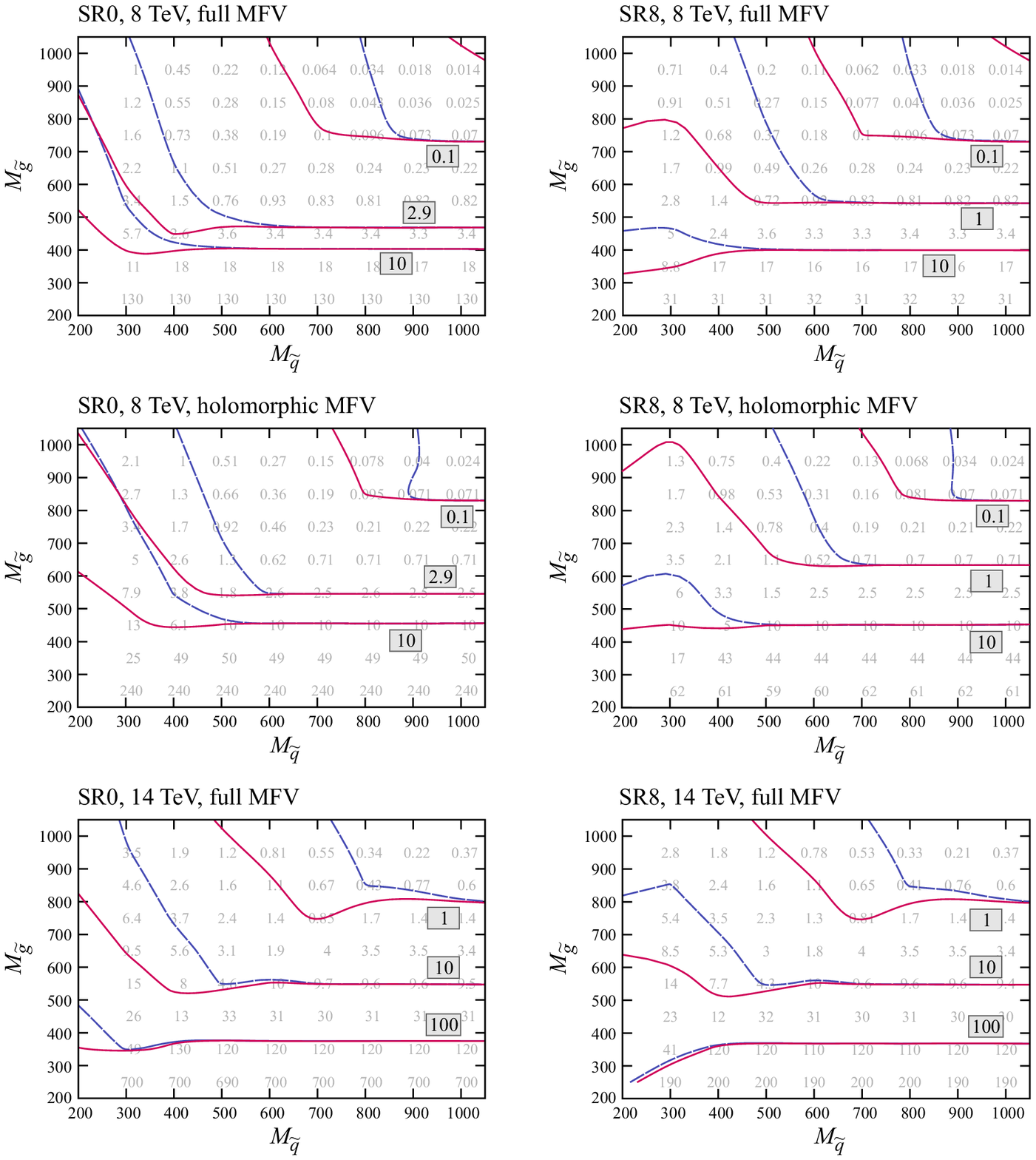}
\caption{Fiducial cross sections [fb] in the SR0 and SR8 signal regions for the same-sign dilepton RPV signal at the LHC. At $8$~TeV, in SR8 (SR0), the $1$~fb ($2.9$~fb) contour line correspond to the $95\%$ CL set by CMS in~\cite{CMS8}. The red (plain) contours are obtained with the contribution of the $\tilde{d}_{R}$ only while those in blue (dashed) assume an equal contribution of $\tilde{d}_{R}$ and $\tilde{d}_{L}$. Comparing with Fig.~\ref{FigBare}, the overall acceptance for the same-sign dilepton RPV signal, including top branching fractions, is between $0.25\%$ and $0.5\%$, comparable to the $(0.29\pm0.04)\%$ quoted by CMS for the SM $t\,\bar{t}$ events~\cite{CMS7,CMS8}.}%
\label{SR0SR8}%
\end{figure}

In the full MFV hierarchy case, we note that signal regions with low $H_{T}$ cuts perform well in the low mass range, where jets are softer.  Everywhere else, SR8 characterized by no $\slashed E_{T}$ cut and a relatively high $H_{T}>320$~GeV requirement provides the best sensitivity. As expected, in the presence of R-parity violation, the SUSY searches requiring a large amount of missing energy are not the best suited. This can be understood from the shapes of the RPV signal and $t\bar tW+t\bar tZ$ background in the $H_{T}-\slashed E_{T}$ plane (see Fig.~\ref{HTET}). For squark and gluino masses close to the exclusion contour of SR8 (without $\tilde d_L$ contributions), the two missing energy distributions are very similar. For higher sparticle masses, the average $\slashed E_{T}$ is only slightly more important in signal events. On the other hand, a relatively good discrimination between signal and background is provided by the transverse hadronic activity $H_{T}$. The jet multiplicity or highest jet $p_{T}$ may also provide powerful handles~\cite{Allanach:2012vj}.

In the whole squark mass range, the SR8 limit excludes gluino masses below roughly $550$~GeV. In the low- and mid-range squark mass region however, the bound varies significantly depending on the contributions of $\tilde{d}_{L}$ to the same-sign tops signal. In the most unfavorable situation where $\tilde{d}_{L}$ contributions are vanishing, the gluino mass limit saturates around $800$~GeV while it rises well above the TeV in the most favorable case where $\tilde{d}_{L}$ contributes as much as $\tilde{d}_{R}$. Note that the same-sign squarks production cross section decreases with increasing gluino masses, so the bound will nonetheless reach a maximum there.

In the holomorphic MFV hierarchy case, the final state $b$ multiplicity is on average higher than with the full MFV hierarchy. Tagging at least two $b$ jets is therefore much more likely and the limits slightly improve. SR7 where three $b$ tags are required is then also populated by a significant number of signal events and provides competitive bounds. Overall, this pushes the limit on sparticle masses higher, towards regions where the average $\slashed E_{T}$ of signal events slightly increases. There, SR3 and SR6 characterized by a higher $\slashed E_{T}>120$~GeV cut and very small backgrounds perform more and more efficiently. This is especially visible when the contributions of $\tilde{d}_{L}$ are significant and further enhance the signal rate. For moderate sparticle masses though, SR8 still leads to the best limit.

We note that our exclusion regions in the holomorphic MFV case are somewhat more conservative than the $M_{\tilde{g}}\gtrsim 800$~GeV limit obtained in Ref.~\cite{BergerPST13}. To see this, first note that the scenario analyzed there decouples all sparticles except the gluino and a top squark, the latter being the LSP. Same-sign top pairs are produced though $p\,p \rightarrow \tilde{g}\, \tilde{g}$ with the gluino decaying as $\tilde{g} \rightarrow t\,b\,s , \bar{t}\,\bar{b}\,\bar{s}$ via on-shell $\tilde{t}$ squarks. As explained in section~\ref{EffDecCh}, such a scenario is covered by our simplified theoretical setting: it corresponds to the $M_{\tilde{q}} \rightarrow \infty$ region of our plots. So, looking at Fig.~\ref{FigLimits}, we get the lower $M_{\tilde{g}}\gtrsim 630$~GeV limit. We checked explicitly that it does not depend significantly on whether the stop can be on-shell or not. Even though the kinematics is different, the selection criteria are broad enough to prevent a significant loss of sensitivity. Now, as can be seen in Fig.~\ref{FigBare}, our LO rate at $M_{\tilde{g}}\approx 800$~GeV is about five times smaller than that at $630$~GeV, where our limit rests. But, as said before, we do not include the NLO corrections. Comparing our Fig.~\ref{FigBare} with Ref.~\cite{ggNLO}, the rate at $800$~GeV is strongly enhanced and nears that computed at LO for $630$~GeV. In addition, there are other subleading but not necessarily negligible differences in the two treatments, for instance: only the $g\,g$ contribution to the gluino pair production has been considered here, the sensitivity is slightly different when stops are on or off their mass-shell, finite-width effects are not included in Ref.~\cite{BergerPST13}, and our simulation procedure is simpler, with for instance the isolated lepton identification and $b$ tag efficiencies kept frozen at $60\%$.

To illustrate the perspectives of improvement on the mass bounds, the fiducial $8$~TeV cross sections for SR8 (currently providing the best sensitivity in most cases) and SR0 (the baseline selection) are displayed in Fig.~\ref{SR0SR8}. Improving the limits by a factor of ten could lead to an increase of the absolute bound on the gluino mass of the order of a couple of hundred GeV. The improvement would be the more significant in the lowest allowed squark mass region where the limit on the gluino mass could increase by more than a factor of two. A similar gain would be obtained at the $14$~TeV LHC if a bound on the BSM same-sign dilepton fiducial rate comparable to the one obtained so far at $8$~TeV is achieved. In this respect, it is worth to stress that the characteristics of the signal change as the sparticles get heavier. With increasing bounds on their masses, the signal regions with significant missing energy should become competitive once adequate techniques are put in place to identify the boosted top quarks (see for instance Ref.~\cite{Boosted}). Though a large fraction of the RPV signal is cut away from these regions, very tight limits can be set there since they are mostly free of backgrounds. 

\begin{figure}[t]
\centering
\includegraphics[width=0.96\textwidth]{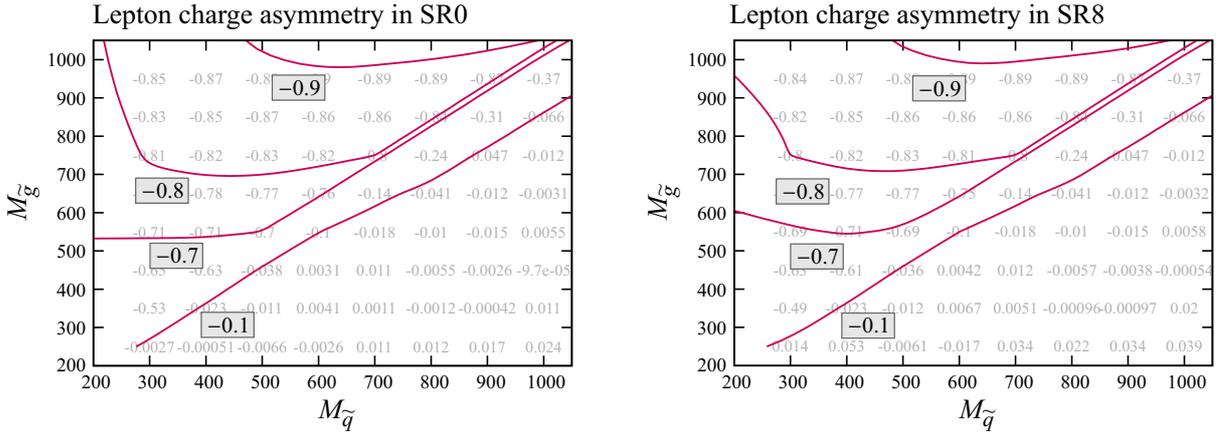}
\caption{Lepton charge asymmetry of Eq.~(\ref{Asym}) exhibited by the same-sign dilepton RPV signal in SR0 and SR8. The $\tilde{d}_{L}$ contributions to the top pair production are not included here.}%
\label{FigAsym}%
\end{figure}

\subsection{Charge asymmetries}

As already mentioned, the irreducible $t\bar{t}\,W$ background features a predominance of positively charged dileptons over negative ones. More quantitatively, {\sc MadGraph5}~\cite{Alwall:2011uj} leading order SM estimates for the lepton charge asymmetry defined in Eq.~(\ref{Asym}) are:%
\begin{equation}
\renewcommand{\arraystretch}{1.4}
\begin{tabular}{r@{\qquad\extracolsep{\fill}}cc@{\myspace}cc@{\myspace}ccc}
 & SR0 & SR1 & SR4 & SR3 & SR8 & SR5 & SR6\\\hline
$A_{\ell\ell^{\prime}}^{t\bar{t}\,W+t\bar tZ} $& 0.26 & 0.28 & 0.29 & 0.36 & 0.26 & 0.28 & 0.35
\end{tabular}\,.
\end{equation}
The value in SR1 agrees well with the central CMS estimate of Ref.~\cite{CMS8}. On the other hand, the RPV processes initiated by down valence quarks (that dominate the same-sign dilepton production when squarks are lighter than gluinos) are significantly more probable than their conjugates, initiated by anti-down quarks. In the upper-left part of the $M_{\tilde{q}}-M_{\tilde{g}}$ plane, much more anti-top than top-quark pairs are therefore expected. This leads to a predominance of negatively charged dileptons and $A_{\ell\ell^{\prime}}$ approaches $-1$ for all $\ell,\ell^{\prime}=e,\mu,\tau$ (see Fig.~\ref{FigAsym}, where only electrons and muons are considered).

This observation has two important consequences. On the theoretical side, as already emphasized in Ref.~\cite{DurieuxGMS12}, such a negative asymmetry is a smoking gun for new physics and an important evidence for baryon number violation. It is indeed almost impossible to obtain in other realistic new physics scenarios. On the experimental side, a precise measurement of this asymmetry, in which systematic uncertainties cancel, could provide important constraints on our model. In addition, a limit on the production rate of negatively charged lepton pairs only, for which SM irreducible backgrounds are smaller, could in principle be used to improve the current bounds in the upper half of the $M_{\tilde{q}}-M_{\tilde{g}}$ plane.

%%%%%%%%%%%%%%%%%%%%%%%%%%%%%%%%%%%%%%%%%%%%%%%%%%%%%%%%%%%%%%%%%%%%
\section{Conclusion}

In this paper, we have analyzed in details the same-sign top-quark pair signature of the MSSM in the presence of R-parity violation. To ensure a sufficiently long proton lifetime, we enforce the MFV hypothesis, which predicts negligible lepton number violating couplings and specific flavor hierarchies for those violating baryon number, $\bm{\lambda}^{\prime\prime IJK}U^{I}D^{J}D^{K}$. In this respect, we have considered both the full MFV prediction~\cite{RPVMFV} as well as its holomorphic restriction~\cite{Grossman}, see Table~\ref{Hierarchies}. Our main results are the followings:

\begin{enumerate}
\item By going through all the possible sparticle decay chains, we showed that the same-sign dilepton signature is a generic feature of the MSSM with R-parity violation. Indeed, independently of the specific MFV implementation, most of the dominant processes lead to same-sign top-quark pairs, because the RPV decays of down-type squarks and gauginos always produce top quarks. By contrast, searches for multijet resonances have a much more restricted reach. Actually, only stop intermediate states have a good probability to lead to final states made only of light-quark jets (provided $\tilde{t}\rightarrow \tilde{g}\,t$ is kinematically closed). 

\item Since the same-sign dilepton signature is to a large extent universal, it can be conveniently simulated using a simplified theoretical framework, thereby avoiding complicated scans over the MSSM parameter space. In practice, it suffices to include only the $g\,g \rightarrow \tilde{g}\,\tilde{g}$, $g\,d \rightarrow \tilde{g}\,\tilde{d}_i$, and $d\,d \rightarrow \tilde{d}_i\,\tilde{d}_j$ ($i,j=L,R$) sparticle production mechanisms, to tune their respective strength by varying the sparticle masses $M_{\tilde{q}}$ and $M_{\tilde{g}}$, and to allow for the sparticle RPV decay through either $\tilde{g} \rightarrow t+2j, \bar{t}+2j$ or $\tilde{d}_i \rightarrow \bar{t}+j$, with only light-quark jets in the full MFV case, or with some $b$ jets in the holomorphic case. A robust estimate of the final limit range for all possible MSSM mass and mixing parameters is obtained by turning completely on and off the contribution of $\tilde d_L$.

\item Using this benchmark strategy, we obtained the approximate exclusion regions shown in Fig.~\ref{FigLimits} from the current CMS dilepton searches, using either the full or holomorphic MFV hierarchies. The bounds are typically tighter for the latter thanks to the more numerous $b$-quark jets. In the future, these exclusion regions are expected to creep upwards. Pushing them well beyond the TeV appears difficult though, and would require new dedicated techniques. In this respect, tailored cuts in transverse missing energy $\slashed E_T$ or hadronic activity ($H_T$, jet multiplicity, jet $p_T$, etc.) as well as information from the lepton charge asymmetry could be exploited. It is also worth to keep in mind that the average hadronic activity, and to a lesser extent the average $\slashed E_T$, increase with sparticle masses. Once the region just above the electroweak scale is cleared, a better sensitivity to the RPV signal could be achievable.

\item  It is well known that sparticles could be rather long-lived even when R-parity is violated. Given the strong suppression of the $\bm{\lambda}^{\prime\prime 1IJ}$, this is especially true for up-type squarks, which could be copiously produced at the LHC. So, we analyzed in details the lifetimes of the squarks, gluino, and to some extent, neutralino and sleptons. We find that except with the holomorphic MFV hierarchy at small $\tan\beta$, sparticles tend to decay rather quickly, see Fig.~\ref{FigLongLife}. This remains true even when the dominant top-producing channels are kinematically closed. Note that the gaugino lifetimes can always be extended by sending squark masses well beyond the TeV scale since their decays proceed through virtual squarks. But, provided squark masses are not too heavy, no viable R-hadron candidates in the $\sim 100$ to $\sim 1000$~GeV mass range are possible once MFV is imposed and $\tan\beta\gtrsim 15$.

\item Neither the stop nor the neutralino are playing an important role in our analysis, because quite independently of their masses, they do not significantly affect the same-sign top-quark pair production rate. So, given the CMS dilepton bounds, these particles could still be very light. If the stop is the LSP, the best strategy to constrain its mass remains to look for a single or a pair of two-jet resonances that would arise from $p\,p\rightarrow \bar{\tilde{t}}+\text{jets}$ or $p\,p\rightarrow \tilde{t}\,\bar{\tilde{t}}+\text{jets}$. For a neutralino LSP, assuming all the other sparticles are far heavier, the same-sign top-pair signal may still be useful, though the signal strength should be rather suppressed since one has to rely on the electroweak interactions to produce pairs of neutralinos. Note, though, that this would not hold if the neutralino becomes long-lived. In the presence of a large MSSM mass hierarchy, and with very suppressed $\bm{\lambda}^{\prime\prime}$ couplings, the best handle would be the search for the monotop signals~\cite{Monotops} produced via $s\,d\rightarrow \bar{\tilde{t}} \rightarrow \bar{t}\,\tilde{\chi}^0$.

\item On a more technical side, we clarified several points concerning squark and gaugino decay rates in the presence of the baryonic RPV couplings. In particular, we observed that the Majorana nature of the gluino (or neutralino) does not always imply the equality of the processes involving their decays into conjugate final states. This is shown analytically for the squark four-body decay processes: $\mathcal{B}(\tilde{q}_{L,R}\rightarrow q\,t\,d\,s)\neq\mathcal{B}(\tilde{q}_{L,R}\rightarrow q\,\bar{t}\,\bar{d}\,\bar{s})$ even though $\mathcal{B}(\tilde{g},\,\tilde{\chi}^0\rightarrow t\,d\,s)=\mathcal{B}(\tilde{g},\,\tilde{\chi}^0\rightarrow\bar{t}\,\bar{d}\,\bar{s})$, see appendix~\ref{App4B}. The reasons for this are the chiral nature of the RPV and gluino couplings, as well as the width of the latter. At leading order, this effect appears to be numerically small for $\sigma(g\,g\rightarrow \tilde{g}\tilde{g}\rightarrow t\,t+\text{jets})$, whose ratio with $\sigma(g\,g\rightarrow \tilde{g}\tilde{g})$ stays close to the expected $1/4$.
\end{enumerate}

In conclusion, though baryonic R-parity violation may appear as a naughty twist of Nature, requiring us to delve into the intense hadronic activity of proton colliders, the LHC may actually be well up to the challenge. First, most of this hadronic activity should be accompanied with top or anti-top quarks, which can be efficiently identified by both CMS and ATLAS. Second, from a baryon number point-of-view, the LHC is an asymmetric machine since it collides protons. This could prove invaluable to disentangle $B$-violating effects from large SM backgrounds. So, even R-parity violating low-scale supersymmetry should not remain unnoticed for long under the onslaught of the future nominal 14~TeV collisions.

\subsubsection*{Acknowledgments}

We would like to thank Sabine Kraml and Fabio Maltoni for discussions and comments. G.D. would like to thank the LPSC, and C.S. the CP3 institute, for their hospitality. G.D. is a Research Fellow of the F.R.S.-FNRS Belgium. This research has been supported in part by the IISN \emph{Fundamental 
interactions}, convention 4.4517.08, and by the Belgian IAP Program BELSPO P7/37.

\pagebreak 

\appendix                      

%%%%%%%%%%%%%%%%%%%%%%%%%%%%%%%%%%%%%%%%%%%%%%%%%%%%%%%%%%%%%%%%%%%%
\section{Decay widths\label{AppDecay}}

The decay widths of squarks, gluinos, and neutralinos in the presence of the R-parity violating couplings $\bm{\lambda}^{\prime\prime}$ have been computed in several places, see in particular Ref.~\cite{Barbier04} and references there. Our purposes here are first to collect (and sometimes correct) the relevant expressions for the two and three body decay processes, $\Gamma(\tilde{q}^{I}\rightarrow\bar{q}^{J}\bar{q}^{K})$ and $\Gamma(\tilde{g},\tilde{\chi}_{1}^{0}\rightarrow q^{I}q^{J}q^{K},\bar{q}^{I}\bar{q}^{J}\bar{q}^{K})$. Second, the four-body squark decay $\tilde{q}^{A}\rightarrow q^{A}q^{I}q^{J}q^{K}$ and $\tilde{q}^{A}\rightarrow q^{A}\bar{q}^{I}\bar{q}^{J}\bar{q}^{K}$ are analyzed and their rates computed. Though significantly phase-space suppressed, hence usually disregarded, these processes become dominant when the $\bm{\lambda}^{\prime\prime}$ couplings able to induce the two-body decays are very suppressed. Finally, as a by-product, we also present the slepton and sneutrino four-body decay rates $\Gamma(\tilde{\ell}^{A}(\tilde{\nu}^{A})\rightarrow\ell^{A}(\nu^{A})q^{I}q^{J}q^{K})$, $\Gamma(\tilde{\ell}^{A}(\tilde{\nu}^{A})\rightarrow\ell^{A}(\nu^{A})\bar{q}^{I}\bar{q}^{J}\bar{q}^{K})$, which would be the only open channels if these particles were the LSP.

\subsection{Two-body squark decays\label{App2B}}

In terms of gauge eigenstates, the two-body decay widths for $\tilde{u}_{R}^{I}\rightarrow \bar{d}^{J}\bar{d}^{K}$ and $\tilde{d}_{R}^{J}\rightarrow \bar{u}^{I}\bar{d}^{K}$ are (Fig.~\ref{FigApp}$a$)%
\begin{align}
\Gamma(\tilde{u}_{R}^{I}\overset{}{\rightarrow}\bar{d}^{J}\bar{d}^{K}) &
=\frac{M_{\tilde{u}^{I}}^{2}-m_{d^{J}}^{2}-m_{d^{K}}^{2}}{8\pi
M_{\tilde{u}^{A}}}\Lambda(\tilde{u}_{R}^{I},d^{J},d^{K})\times|%
\bm{\lambda}_{IJK}^{\prime\prime}|^{2}\;,\\
\Gamma(\tilde{d}_{R}^{J}\overset{}{\rightarrow}\bar{u}^{I}\bar{d}^{K}) &  =\frac{M_{\tilde
{d}^{J}}^{2}-m_{u^{I}}^{2}-m_{d^{K}}^{2}}{8\pi M_{\tilde{d}^{A}}}%
\Lambda(\tilde{d}_{R}^{J},u^{I},d^{K})\times|\bm{\lambda}_{IJK}^{\prime\prime}|^{2}\;,
\end{align}
while $\Gamma(\tilde{u}_{L}^{I}\rightarrow \bar{d}^{J}\bar{d}^{K})=\Gamma(\tilde{d}_{R}^{J}\rightarrow \bar{u}^{I}\bar{d}^{K})=0$. The standard kinematical function is $\Lambda(a,b,c)=\lambda(1,m_{b}^{2}/m_{a}^{2},m_{c}^{2}/m_{a}^{2})$ with $\lambda(a,b,c)^{2}=a^{2}+b^{2}+c^{2}-2(ab+ac+bc)$.

These gauge eigenstates mix into mass eigenstates. Introducing the $6\times6$ mixing matrices $H^{f}$, $f=u,d,e$, relating the mass eigenstates $\tilde{f}^{A}$, $A=1,...,6$ to the gauge eigenstates $(\tilde{f}_{L}^{I},\tilde{f}_{R}^{I})$, $I=1,2,3$, the rates become
\begin{align}
\Gamma(\tilde{d}^{A}\overset{}{\rightarrow}\bar{u}^{I}\bar{d}^{K})  &  =\frac{M_{\tilde{d}^{J}}^{2}-m_{u^{I}}^{2}-m_{d^{K}}^{2}}{8\pi M_{\tilde{d}^{A}}}
\Lambda(\tilde{d}^{A},u^{I},d^{K})\times|\bm{\lambda}_{ILK}^{\prime\prime}H_{A(L+3)}^{d}|^{2}\;,\\
\Gamma(\tilde{u}^{A}\overset{}{\rightarrow}\bar{d}^{J}\bar{d}^{K})  &
=\frac{M_{\tilde{u}^{I}}^{2}-m_{d^{J}}^{2}-m_{d^{K}}^{2}}{8\pi M_{\tilde{u}^{A}}}\Lambda(\tilde{u}^{A},d^{J},d^{K})\times|\bm{\lambda}_{LJK}^{\prime\prime}H_{A(L+3)}^{u}|^{2}\;.
\end{align}
Under MFV, the four blocks $H_{IJ}^{f},\;H_{(I+3)(J+3)}^{f}$, $H_{I(J+3)}^{f}$ and $H_{(I+3)J}^{f}$, $I,J=1,2,3$, are close to diagonal (exactly diagonal for $f=e$). When flavor mixings are neglected, we define a separate LR mixing matrix for each squark and slepton flavor, so that
\begin{equation}
\Theta^{f^{I}}\equiv\left(
\begin{array}[c]{cc}%
H_{II}^{f} & H_{I(I+3)}^{f}\\
H_{(I+3)I}^{f} & H_{(I+3)(I+3)}^{f}%
\end{array}
\right)  \rightarrow\Gamma(\tilde{f}_{i}\rightarrow X)=|\Theta_{iL}^{f}%
|^{2}\times\Gamma(\tilde{f}_{L}\rightarrow X)+|\Theta_{iR}^{f}|^{2}%
\times\Gamma(\tilde{f}_{R}\rightarrow X)\;. \label{USM}%
\end{equation}
For example, when only $\bm{\lambda}_{tds}^{\prime\prime}$ is significant (and using $\bm{\lambda}_{tds}^{\prime\prime}=-\bm{\lambda}_{tsd}^{\prime\prime\ast}$), the allowed two-body decay channels are%
\begin{align}
\Gamma(\tilde{d}_{i}\overset{}{\rightarrow}\bar{t}\,\bar{s})  &  \approx(13\;\text{GeV)}%
\times| \bm{\lambda}_{tds}^{\prime\prime}|^{2}\times|\Theta_{iR}^{d}|^{2}\;,\\
\Gamma(\tilde{s}_{i}\overset{}{\rightarrow}\bar{t}\,\bar{d})  &  \approx(13\;\text{GeV)}%
\times|\bm{\lambda}_{tds}^{\prime\prime}|^{2}\times|\Theta_{iR}^{s}|^{2}\;,\\
\Gamma(\tilde{t}_{i}\overset{}{\rightarrow}\bar{d}\,\bar{s})  &  \approx(18\;\text{GeV)}%
\times|\bm{\lambda}_{tds}^{\prime\prime}|^{2}\times|\Theta_{iR}^{t}|^{2}\text{\ },
\end{align}
for squark masses of 450 GeV. Note that under MFV, the LR mixings are tuned by the quark masses, so $\Theta_{1R}^{s,d}\ll\Theta_{2R}^{s,d}\approx1$.

\subsection{Three-body gaugino decays\label{App3B}}

\begin{figure}[t]
\centering
\includegraphics[width=15.5cm]{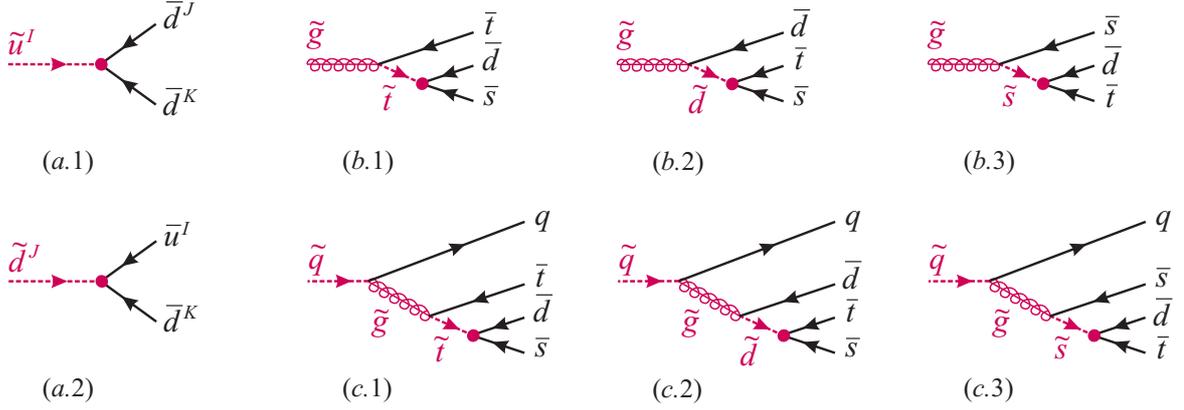}
\caption{The squark two-body ($a$), gluino three-body ($b$), and squark four-body decay processes ($c$) induced by the RPV couplings $\bm{\lambda}^{\prime\prime}$. For ($b$) and ($c$), the diagrams with $t\,d\,s$ instead of $\bar{t}\,\bar{d}\,\bar{s}$ or with a neutralino instead of a gluino are similar. For the four-body decays, crossed diagrams are understood when $q$ is identical to one of the other quarks in the final state.}%
\label{FigApp}%
\end{figure}

When light, the gluino and lightest neutralino decay predominantly through virtual squark exchanges, see Fig.~\ref{FigApp}$b$. The amplitudes and decay rates in the general case are rather involved, so we introduce a few approximations. First, we keep only one RPV coupling as significant, and take $\bm{\lambda}_{tds}^{\prime\prime}$ for definiteness. Second, up squarks are considered degenerate in mass, and so are down squarks. From this, the sum over the virtual squark six states simplifies thanks to the unitarity of the squark mixing matrices (GIM mechanism). Third, this also implies that the wino contribution cancels out, leaving only the bino and Higgsinos. Since the latter couplings are tuned by the quark Yukawa couplings, we consider only the bino component of $\tilde{\chi}_{1}^{0}$ in the following.

Under these simplifications, the decay amplitudes take the form
\begin{subequations}
\label{Ampli3B}%
\begin{align}
\mathcal{M}(\lambda &  \rightarrow\bar{t}\,\bar{d}\,\bar{s})=g_{tsd}^{\lambda
}\{\bar{v}_{\lambda}P_{R}v_{t}\}\{\bar{u}_{s}P_{R}v_{d}\}+g_{std}^{\lambda
}\{\bar{v}_{\lambda}P_{R}v_{s}\}\{\bar{u}_{t}P_{R}v_{d}\}+g_{dts}^{\lambda
}\{\bar{v}_{\lambda}P_{R}v_{d}\}\{\bar{u}_{t}P_{R}v_{s}\}\;,\\
\mathcal{M}(\lambda &  \rightarrow t\,d\,s)=g_{tsd}^{\lambda\ast}\{\bar
{v}_{\lambda}P_{L}v_{t}\}\{\bar{u}_{s}P_{L}v_{d}\}+g_{std}^{\lambda\ast}%
\{\bar{v}_{\lambda}P_{L}v_{s}\}\{\bar{u}_{t}P_{L}v_{d}\}+g_{dts}^{\lambda\ast
}\{\bar{v}_{\lambda}P_{L}v_{d}\}\{\bar{u}_{t}P_{L}v_{s}\}\;,
\end{align}
with $P_{L,R}=(1\mp\gamma_{5})/2$ and%
\end{subequations}
\begin{equation}
g_{abc}^{\tilde{g}^{\alpha}}=-\bm{\lambda}_{tds}^{\prime\prime\ast}\sqrt{2}g_{S}\frac{\varepsilon^{c_{b}c_{c}c_{d}}T_{c_{d}c_{a}}^{\alpha}}{(p_{b}+p_{c})^{2}-M_{\tilde{a}}^{2}}\;,\;\;g_{abc}%
^{\tilde{\chi}_{1}^{0}}=-\bm{\lambda}_{tds}^{\prime\prime\ast}\frac{Y_{a_{R}}eN_{1B}}{\sqrt{2}\cos\theta_{W}}\frac{\varepsilon^{c_{a}c_{b}c_{c}}}{(p_{b}+p_{c})^{2}-M_{\tilde{a}}^{2}}\;,
\label{ggauginos}
\end{equation}
where $T_{ij}^{\alpha}$ are $SU(3)_{C}$ generators, $\alpha$ is an adjoint color index, $c_{a,b,c,d}$ are fundamental color indices (summation over repeated indices is understood), $g_{S}$ and $e$ are the strong and electromagnetic coupling constants, $\theta_{W}$ is the Weinberg angle, $Y_{a_{R}}$ is the weak hypercharge of $a_{R}$ ($Y_{t_{R}}=4/3$ and $Y_{d_{R}}=Y_{s_{R}}=-2/3$), and $N_{1B}$ is the mixing angle between the bino gauge eigenstate and the lightest neutralino mass eigenstate. Under conjugation $g\rightarrow g^{\ast}$, it is understood that $\bm{\lambda}_{tds}^{\prime\prime\ast}\rightarrow\bm{\lambda}_{tds}^{\prime\prime}$, $N_{1B}\rightarrow N_{1B}^{\ast}$, and $T_{ij}^{\alpha}\rightarrow T_{ji}^{\alpha}$, but $Y_{a}$ and $\varepsilon^{c_{a}c_{b}c_{c}}$ stay put. In Eq.~(\ref{ggauginos}), we set the widths of the squarks to zero in their respective propagators since we are only interested in the situation where they are relatively far off their mass shell.

The squared amplitudes have to be summed over the quark spins and color indices, and averaged over the gaugino spins as well as, and in the gluino case, over the adjoint color index. The sum over the spins can be done using the usual formulas provided some fermion lines are inverted using charge conjugation. Then, the squared amplitudes are the same for $\lambda\rightarrow\bar{t}\,\bar{d}\,\bar{s}$ and $\lambda\rightarrow t\,d\,s$,
\begin{align}
|\mathcal{M}(\lambda\overset{}{\rightarrow}t\,d\,s)|^{2}\overset{}{=}%
|\mathcal{M}(\lambda\overset{}{\rightarrow}\bar{t}\,\bar{d}\,\bar{s})|^{2} &
=4|g_{tsd}|^{2}p_{\lambda}\cdot p_{t}\,p_{s}\cdot p_{d}+4\operatorname{Re}%
(g_{tsd}^{\ast}g_{std})g(p_{\lambda},p_{t},p_{s},p_{d})\nonumber\\
&  \;+4|g_{std}|^{2}p_{\lambda}\cdot p_{s}\,p_{t}\cdot p_{d}-4\operatorname{Re}%
(g_{tsd}^{\ast}g_{dts})g(p_{\lambda},p_{t},p_{d},p_{s})\nonumber\\
&  \;+4|g_{dts}|^{2}p_{\lambda}\cdot p_{d}\,p_{t}\cdot p_{s}+4\operatorname{Re}%
(g_{dts}^{\ast}g_{std})g(p_{\lambda},p_{d},p_{s},p_{t})\;,
\end{align}
with $g(a,b,c,d)=(a\cdot b)(c\cdot d)+(a\cdot c)(b\cdot d)-(a\cdot d)(b\cdot c)$. Summation over the color indices is understood for the $g_{abc}g_{def}^{\ast}$ coefficients, and can be done using the standard formulas:%
\begin{equation}
\varepsilon^{ijk}\varepsilon^{lmn}=\det\left(
\begin{array}[c]{ccc}%
\delta^{il} & \delta^{im} & \delta^{in}\\
\delta^{jl} & \delta^{jm} & \delta^{jn}\\
\delta^{kl} & \delta^{km} & \delta^{kn}%
\end{array}
\right)  \;,\;\;\sum_{a=1}^{8}T_{ij}^{a}T_{kl}^{a}=\frac{1}{2}(\delta_{il}\delta_{jk}-\frac{1}{3}\delta_{ij}\delta_{kl})\;.
\end{equation}
From the squared amplitudes, the gaugino RPV decay rates are%
\begin{equation}
\Gamma(\lambda\rightarrow t\,d\,s)=\Gamma(\lambda\rightarrow\bar{t}\,\bar{d}\,\bar
{s})=\frac{1}{(2\pi)^{3}}\frac{C_{\lambda}}{32M_{\lambda}^{3}}\int
d\Phi_{\lambda\rightarrow tds}\sum_{spins}|\mathcal{M}(\lambda\rightarrow
tds)|^{2}\;,
\end{equation}
with $C_{\tilde{\chi}_{1}^{0}}=1/2$ and $C_{\tilde{g}}=1/2\times1/8$ for the spin and color averages. For the neutralino case, we reproduce the result of Ref.~\cite{BaltzG97} once the GIM mechanism is enforced and non-bino contributions discarded. As noted there, this result disagrees with the earlier computation done in Ref.~\cite{ButterworthD92}, in which the interference terms appear to drop out in the massless quark limit (the same holds for Ref.~\cite{DreinerRS99}, quoted in Ref.~\cite{Barbier04}). By contrast, we find that for both the neutralino and gluino decays, the interference terms survive in that limit.

\begin{figure}[t]
\centering
\includegraphics[width=0.96\textwidth]{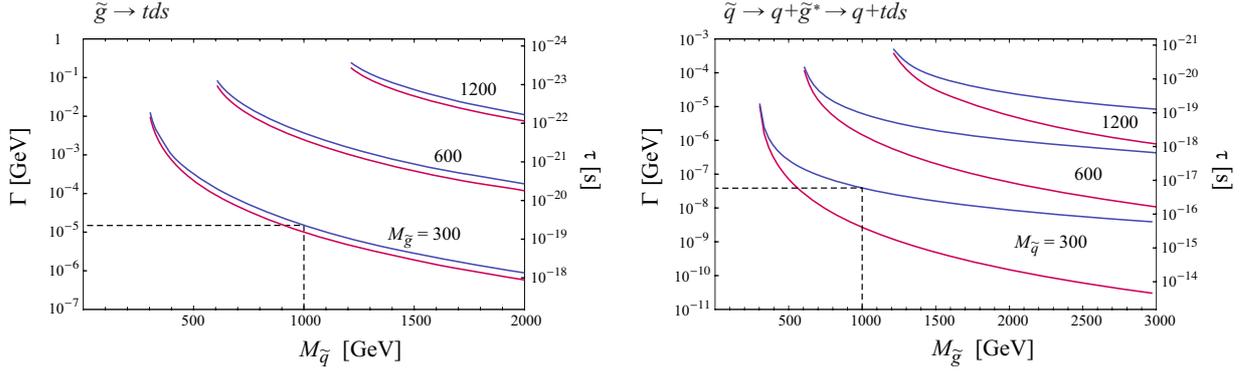}
\caption{Left: The gluino partial decay width and lifetime, for $\bm{\lambda}_{tds}^{\prime\prime}= 1$, as a function of the virtual squark mass $M_{\tilde{q}}=M_{\tilde{t}}=M_{\tilde{d}}=M_{\tilde{s}}$. The lower (red) curves show the impact of neglecting interference terms: after a slight increase, it quickly settles at its asymptotic value of $3/2$. Right: typical four-body squark partial decay rate and lifetime, again for $\bm{\lambda}_{tds}^{\prime\prime} = 1$ and degenerate squarks, as a function of the virtual gluino mass. This time, interference terms are neglected. The lower (red) curves show the rate for the $\mathcal{I}^{D},\mathcal{J}^{D}$ contributions, and the upper (blue) ones the $\mathcal{I}^{M},\mathcal{J}^{M}$ contributions. The much slower decoupling of the latter is due to the additional factor of $M_{\tilde{g}}$ required for the chirality flip. In both figures, the corresponding rates for neutralinos can be obtained by a simple rescaling, see Eqs.~(\ref{gluneuser}) and (\ref{gluneuGamma}). Finally, the values of the rates when the mass of the virtual squark or gluino is 1 TeV correspond to those quoted in section~3.}%
\label{FigAppLT}%
\end{figure}

The phase space measure $d\Phi_{\lambda\rightarrow tds}$ can be written in terms of the usual Dalitz plot variables $m_{ab}^{2}=(p_{a}+p_{b})^{2}$. In the limit where $m_{d},m_{s}\rightarrow0$, the integration limits are rather simple,%
\begin{equation}
\int d\Phi_{\lambda\rightarrow tds}=\int_{m_{t}^{2}}^{M_{\lambda}^{2}}dm_{ts}^{2}\int_{0}^{(M_{\lambda}^{2}-m_{ts}^{2})(m_{ts}^{2}-m_{t}^{2})/m_{ts}^{2}}dm_{sd}^{2}\;.
\end{equation}
Even setting $m_{t}$ to zero and taking all squarks degenerate (with mass $M_{\tilde{q}}$), the analytic expression for the fully integrated rate is quite complicated. In the $M_{\tilde{q}}/M_{\lambda}\rightarrow\infty$ limit, both $g_{abc}^{\tilde{g}^{\alpha}}$ and $g_{abc}^{\tilde{\chi}_{1}^{0}}$ become momentum independent, and the differential rates are easily integrated:
\begin{subequations}
\label{gluneuser}
\label{RateSeries}%
\begin{align}
\Gamma(\tilde{g} &  \rightarrow t\,d\,s)=\frac{\alpha_{S}M_{\tilde{g}}}{256\pi^{2}}\times|\bm{\lambda}_{tds}^{\prime\prime}|^{2}\times\frac{M_{\tilde{g}}^{4}}{M_{\tilde{q}}^{4}}\times\left(  1+\frac{1}{2}\right)  +\mathcal{O}\left( \frac{M_{\tilde{g}}^{6}}{M_{\tilde{q}}^{6}}\right)  \;,\\
\Gamma(\tilde{\chi}_{1}^{0} &  \rightarrow t\,d\,s)=\frac{\alpha M_{\tilde{\chi}^{0}} |N_{1B}|^2}{192\pi^{2}\cos^{2}\theta_{W}}\times|\bm{\lambda}_{tds}^{\prime\prime}|^{2}\times\frac{M_{\tilde{\chi}_{1}^{0}}^{4}}{M_{\tilde{q}}^{4}}\times\left( 1+\frac{1}{2}\right) +\mathcal{O}\left( \frac{M_{\tilde{\chi}_{1}^{0}}^{6}}{M_{\tilde{q}}^{6}}\right)  \;,
\end{align}
where the $1/2$ in the final brackets originate from the interference terms. The fact that both amount to a $50\%$ correction is coincidental.

Note that these expressions are not to be used when the gaugino and squark masses are close, or when the gaugino is not sufficiently heavy to justify setting the top-quark mass to zero. In these cases, the phase-space integrals have to be performed numerically (we actually rely on the \textsc{FeynRules}--\textsc{MadGraph5} software chain~\cite{Christensen:2008py, Alwall:2011uj} for our simulations). For example, taking $M_{\lambda}=450$~GeV, $M_{\tilde{q}}=600$~GeV, $\alpha_{S}=0.1$, $\alpha=1/128$, $|N_{1B}|=1$, and $m_{t}=170$~GeV gives%
\end{subequations}
\begin{align}
\Gamma(\tilde{g} &  \rightarrow t\,d\,s)=|\bm{\lambda}_{tds}^{\prime\prime}|^{2}\times(3.7+1.8)\times10^{-3}\;\text{GeV}\;,\;\\
\Gamma(\tilde{\chi}_{1}^{0} &  \rightarrow t\,d\,s)=|\bm{\lambda}_{tds}^{\prime\prime}|^{2}\times(4.3+2.1)\times10^{-4}\;\text{GeV}\;,
\end{align}
where the first (second) numbers in the brackets denote the direct (interfering) contributions. For comparison, Eq.~(\ref{RateSeries}) give the slightly larger estimates $\Gamma(\tilde{g}\rightarrow t\,d\,s)=|\bm{\lambda}_{tds}^{\prime\prime}|^{2}\times6.6\times10^{-3}\;$GeV and $\Gamma(\tilde
{\chi}_{1}^{0}\rightarrow t\,d\,s)=|\bm{\lambda}_{tds}^{\prime\prime}|^{2}\times6.8\times10^{-4}\;$GeV. Finally, the evolution of the gluino lifetime as a function of its mass as well as that of the virtual squarks is shown in Fig.~\ref{FigAppLT}.

\subsection{Four-body squark decays\label{App4B}}

The four-body processes shown in Fig.~\ref{FigApp}$c$ are relevant when there is a large flavor hierarchy between the RPV couplings. Indeed, when the two-body decay is very suppressed, it becomes advantageous to proceed through a virtual gluino or neutralino which then decays via the largest RPV coupling. Under the same simplifying assumptions as for the gluino and neutralino decays, the amplitudes can be obtained from Eq.~(\ref{Ampli3B}) as
\begin{subequations}
\label{Ampli4B}%
\begin{align}
\mathcal{M}(\tilde{q}_{i}\overset{}{\rightarrow}q\,\tilde{g}^{\alpha}\overset
{}{\rightarrow}q\,X) &  =\bar{u}_{q}\left(  \Theta_{i2}^{q\ast}P_{L}-\Theta
_{i1}^{q\ast}P_{R}\right)  \frac{\sqrt{2}g_{S}T_{c_{q}c_{\tilde{u}}}^{\alpha}%
}{\slashed p_{\tilde{g}}+M_{\tilde{g}}}\mathcal{\hat{M}}(\tilde{g}^{\alpha
}\rightarrow X)\;,\\
\mathcal{M}(\tilde{q}_{i}\overset{}{\rightarrow}q\,\tilde{\chi}_{1}^{0}\overset
{}{\rightarrow}q\,X) &  =\bar{u}_{q}\left(  Y_{q_{R}}N_{1B}^{\ast}\Theta
_{i2}^{q\ast}P_{L}-Y_{q_{L}}N_{1B}\Theta_{i1}^{q\ast}P_{R}\right)
\frac{e/(\sqrt{2}\cos\theta_{W})}{\slashed p_{\tilde{\chi}_{1}^{0}}+M_{\tilde{\chi}_{1}^{0}}%
}\mathcal{\hat{M}}(\tilde{\chi}_{1}^{0}\rightarrow X)\;,
\end{align}
where $X=\bar{t}\,\bar{d}\,\bar{s}$ or $t\,d\,s$ and $\mathcal{M}(\lambda\rightarrow X)=\bar{v}_{\lambda}\mathcal{\hat{M}}(\lambda\rightarrow X)$. The two-by-two squark mixing matrices $\Theta^{q}$ are defined in Eq.~(\ref{USM}). Note that for $q=d,t,s$ and $X=t\,d\,s$, one should also include the crossed processes since there are two identical quarks in the final states. We will ignore this complication in the following.

The calculation of the squared amplitudes, summed over spins and colors, proceeds as before, but the four-body phase-space integral cannot be done analytically. As before, we rely on the \textsc{FeynRules}--\textsc{MadGraph5} software chain~\cite{Christensen:2008py, Alwall:2011uj} for our simulations. Still, it is interesting to push the analytical study a bit further, and derive the scaling of the decay rates in terms of the gaugino and virtual squark masses. This is not so trivial since the virtual squarks have masses similar to the initial decaying squark, and thus the momentum dependences of their propagators cannot be neglected. So, to proceed and partly perform the phase-space integrals, we neglect all the interference terms. In the previous section, those were found to increase the gaugino decay rates by $50\%$, so the present computation should not be expected to hold to better than a factor of about two. The three direct contributions can be integrated recursively, leading to
\end{subequations}
\begin{subequations}
\label{Rate4B}%
\begin{align}
\frac{\Gamma(\tilde{q}_{L}\overset{}{\rightarrow}q\,t\,d\,s)_{\tilde{g}}^{dir}%
}{\Gamma_{\tilde{g}}^{0}} &  =\frac{\Gamma(\tilde{q}_{R}\overset{}%
{\rightarrow}q\,\bar{t}\,\bar{d}\,\bar{s})_{\tilde{g}}^{dir}}{\Gamma_{\tilde{g}}%
^{0}}=\mathcal{I}_{\tilde{q},\tilde{t},\tilde{g}}^{D}+\mathcal{J}_{\tilde
{q},\tilde{s},\tilde{g}}^{D}+\mathcal{J}_{\tilde{q},\tilde{d},\tilde{g}}%
^{D}\;,\smallskip\\
\frac{\Gamma(\tilde{q}_{R}\overset{}{\rightarrow}q\,t\,d\,s)_{\tilde{g}}^{dir}%
}{\Gamma_{\tilde{g}}^{0}} &  =\frac{\Gamma(\tilde{q}_{L}\overset{}%
{\rightarrow}q\,\bar{t}\,\bar{d}\,\bar{s})_{\tilde{g}}^{dir}}{\Gamma_{\tilde{g}}%
^{0}}=\mathcal{I}_{\tilde{q},\tilde{t},\tilde{g}}^{M}+\mathcal{J}_{\tilde
{q},\tilde{s},\tilde{g}}^{M}+\mathcal{J}_{\tilde{q},\tilde{d},\tilde{g}}%
^{M}\;,\smallskip\\
\frac{\Gamma(\tilde{q}_{L}\overset{}{\rightarrow}f\,t\,d\,s)_{\tilde{\chi}_{1}^{0}%
}^{dir}}{Y_{q_{L}}^{2}\Gamma_{\tilde{\chi}}^{0}} &  =\frac{\Gamma(\tilde
{q}_{R}\overset{}{\rightarrow}f\,\bar{t}\,\bar{d}\,\bar{s})_{\tilde{\chi}_{1}^{0}%
}^{dir}}{Y_{q_{R}}^{2}\Gamma_{\tilde{\chi}}^{0}}=Y_{u_{R}}^{2}\mathcal{I}%
_{\tilde{q},\tilde{t},\tilde{\chi}}^{D}+Y_{d_{R}}^{2}\mathcal{J}_{\tilde
{q},\tilde{s},\tilde{\chi}}^{D}+Y_{s_{R}}^{2}\mathcal{J}_{\tilde{q},\tilde
{d},\tilde{\chi}}^{D}\;,\smallskip\\
\frac{\Gamma(\tilde{q}_{R}\overset{}{\rightarrow}f\,t\,d\,s)_{\tilde{\chi}_{1}^{0}%
}^{dir}}{Y_{q_{R}}^{2}\Gamma_{\tilde{\chi}}^{0}} &  =\frac{\Gamma(\tilde
{q}_{L}\overset{}{\rightarrow}f\,\bar{t}\,\bar{d}\,\bar{s})_{\tilde{\chi}_{1}^{0}%
}^{dir}}{Y_{q_{L}}^{2}\Gamma_{\tilde{\chi}}^{0}}=Y_{u_{R}}^{2}\mathcal{I}%
_{\tilde{q},\tilde{t},\tilde{\chi}}^{M}+Y_{d_{R}}^{2}\mathcal{J}_{\tilde
{q},\tilde{s},\tilde{\chi}}^{M}+Y_{s_{R}}^{2}\mathcal{J}_{\tilde{q},\tilde
{d},\tilde{\chi}}^{M}\;,
\end{align}
where $Y_{q}$ is the hypercharge of the quark $q$ (remember that under our approximation, the wino and Higgsinos do not contribute), the overall coefficients are%
\end{subequations}
\begin{subequations}
\label{gluneuGamma}
\begin{align}
\Gamma_{\tilde{g}}^{0} &  =\frac{\alpha_{S}^{2}M_{\tilde{q}}}{96\pi^{3}}|\bm{\lambda}_{tds}^{\prime\prime}|^{2}\approx(1\times10^{-3}\,\text{GeV})\times|\bm{\lambda}_{tds}^{\prime\prime}|^{2}\times\frac{M_{\tilde{q}}}{300\,\text{GeV}}\;,\;\;\\
\Gamma_{\tilde{\chi}}^{0} &  =\frac{3\alpha^{2}|N_{1B}|^{4}M_{\tilde{q}}}{1024\pi^{3}\cos^{4}\theta_{W}}|\bm{\lambda}_{tds}^{\prime\prime}|^{2}\approx(3\times10^{-6}\,\text{GeV})\times|\bm{\lambda}_{tds}^{\prime\prime}|^{2}\times\frac{M_{\tilde{q}}}{300\,\text{GeV}}\times
|N_{1B}|^{4}\;,
\end{align}
\end{subequations}
and the dimensionless phase-space integrals can be expressed for $m_{q,d,s}=0$ as
\begin{align}
\left\{
\begin{array}[c]{c}%
\mathcal{I}_{\tilde{q},\tilde{t},\lambda}^{D}\smallskip\\
\mathcal{I}_{\tilde{q},\tilde{t},\lambda}^{M}%
\end{array}
\right.   &  =\int_{m_{t}^{2}}^{M_{\tilde{q}}^{2}}\frac{dT_{\lambda}^{2}}{M_{\tilde{q}}^{2}}\int_{0}^{(T_{\lambda}-m_{t})^{2}}\frac{dT_{\tilde{t}}^{2}}{M_{\tilde{q}}^{2}}\frac{T_{\tilde{t}}^{2}(M_{\tilde{q}}^{2}-T_{\lambda}^{2})^{2}(T_{\lambda}^{2}-T_{\tilde{t}}^{2}+m_{t}^{2})\lambda(T_{\lambda}^{2},T_{\tilde{t}}^{2},m_{t}^{2})}{T_{\lambda}^{4}(T_{\tilde{t}}^{2}-M_{\tilde{t}}^{2})^{2}(T_{\lambda}^{2}-M_{\lambda}^{2})^{2}}\left\{
\begin{array}[c]{c}%
T_{\lambda}^{2}\smallskip\\ M_{\lambda}^{2}%
\end{array}
\right.  \;,\\
\left\{
\begin{array}[c]{c}
\mathcal{J}_{\tilde{q},a,\lambda}^{D}\smallskip\\
\mathcal{J}_{\tilde{q},a,\lambda}^{M}%
\end{array}
\right.   &  =\int_{m_{t}^{2}}^{M_{\tilde{q}}^{2}}\frac{dT_{\lambda}^{2}%
}{M_{\tilde{q}}^{2}}\int_{m_{t}^{2}}^{T_{\lambda}^{2}}\frac{dT_{a}^{2}%
}{M_{\tilde{q}}^{2}}\;\frac{(T_{a}^{2}-m_{t}^{2})^{2}(M_{\tilde{q}}%
^{2}-T_{\lambda}^{2})^{2}(T_{\lambda}^{2}-T_{a}^{2})^{2}}{T_{a}^{2}T_{\lambda
}^{4}(T_{a}^{2}-M_{a}^{2})^{2}(T_{\lambda}^{2}-M_{\lambda}^{2})^{2}}\left\{
\begin{array}[c]{c}T_{\lambda}^{2}\smallskip\\M_{\lambda}^{2}\end{array}
\right.  \;,
\end{align}
and $\lambda(a,b,c)^{2}=a^{2}+b^{2}+c^{2}-2(ab+ac+bc)$. The subscript ``$dir"$ serves as a reminder that interference terms originating from crossed processes when $q=t,d,s$, $X=t\,d\,s$ and from squaring the amplitude are both neglected. Decay rates into mass eigenstates are found using Eq.~(\ref{USM}).

These expressions remain valid if the gluino or the squark in the decay chain can be on-shell, provided their widths are introduced in the denominators of $\mathcal{I}$ and $\mathcal{J}$. In this respect, it is interesting to note that while $\mathcal{B}(\lambda\rightarrow t\,d\,s)=\mathcal{B}(\lambda\rightarrow\bar{t}\,\bar{d}\,\bar{s})$, we find that $\mathcal{B}(\tilde{q}_{L,R}\rightarrow q\,t\,d\,s)\neq\mathcal{B}(\tilde{q}_{L,R}\rightarrow q\,\bar{t}\,\bar{d}\,\bar{s})$ because $\mathcal{I}^{D}\neq\mathcal{I}^{M}$ and $\mathcal{J}^{D}\neq\mathcal{J}^{M}$. This difference can be traced back to the chiral nature of the gaugino-squark-quark and RPV couplings. The projectors in Eq.~(\ref{Ampli4B}) leave only either the $\slashed  p_{\lambda}$ or the $M_{\lambda}$ term of the gaugino propagator to contribute. Because of this, the naive expectation based on the narrow-width approximation should not always be trusted~\cite{Berdine07} (see also Ref.~\cite{DiracGlue,Kramer09}). Numerically, the difference is negligible over most of the $0 < M_{\lambda} < M_{\tilde{q}}$ range when the gaugino width is small, but gets maximal in the deep virtual (massless) limits: $\mathcal{J}^{D}/\mathcal{J}^{M}\rightarrow 0 \,(\infty)$ when $M_{\lambda}\rightarrow \infty \,(0)$.

Specifically, setting all squark masses to a common value $M_{\tilde{q}}$, the phase-space integrals of each type are identical when $m_{t}\rightarrow0$. When $M_{\lambda}\rightarrow 0$, independently of its width,
\begin{align}
\left.  \mathcal{I}_{\tilde{q},\tilde{t},\lambda}^{D}\right|  _{M_{\lambda}\rightarrow 0}  &  =\left.  \mathcal{J}_{\tilde{q},\tilde{s},\lambda}^{D}\right|  _{M_{\lambda}\rightarrow 0}=\left.  \mathcal{J}_{\tilde{q},\tilde{d},\lambda}^{D}\right|  _{M_{\lambda}\rightarrow 0}=\frac{79-8\pi^{2}}{4} \approx0.011\;,\\
\left.  \mathcal{I}_{\tilde{q},\tilde{t},\lambda}^{M}\right|  _{M_{\lambda}\rightarrow 0}  &  =\left.  \mathcal{J}_{\tilde{q},\tilde{s},\lambda}^{M}\right|  _{M_{\lambda}\rightarrow 0}=\left.  \mathcal{J}_{\tilde{q},\tilde{d},\lambda}^{M}\right|  _{M_{\lambda}\rightarrow 0}=0\;.
\end{align}
At the threshold $M_{\lambda}=M_{\tilde{q}}$, the mass-dependent contribution slightly surpasses that of the direct contribution, 
\begin{align}
\left.  \mathcal{I}_{\tilde{q},\tilde{t},\lambda}^{D}\right|  _{M_{\lambda}\rightarrow M_{\tilde{q}}}  &  =\left.  \mathcal{J}_{\tilde{q},\tilde{s},\lambda}^{D}\right|  _{M_{\lambda}\rightarrow M_{\tilde{q}}}=\left.  \mathcal{J}_{\tilde{q},\tilde{d},\lambda}^{D}\right|  _{M_{\lambda}\rightarrow M_{\tilde{q}}%
}=\frac{10-\pi^{2}}{2}\approx0.065\;,\\
\left.  \mathcal{I}_{\tilde{q},\tilde{t},\lambda}^{M}\right|  _{M_{\lambda}\rightarrow M_{\tilde{q}}}  &  =\left.  \mathcal{J}_{\tilde{q},\tilde{s},\lambda}^{M}\right|  _{M_{\lambda}\rightarrow M_{\tilde{q}}}=\left.  \mathcal{J}_{\tilde{q},\tilde{d},\lambda}^{M}\right|  _{M_{\lambda}\rightarrow M_{\tilde{q}}}%
=\frac{4\pi^{2}-39}{6}\approx0.080\;,
\end{align}
while moving into the virtual gaugino regime, the direct contribution rapidly decouples, as can be see expanding the integrals in powers of $M_{\tilde{q}}/M_{\lambda}$ (see Fig.~\ref{FigAppLT}):
\begin{align}
\left.  \mathcal{I}_{\tilde{q},\tilde{t},\lambda}^{D}\right|  _{M_{\lambda}\rightarrow\infty}  &  =\left.  \mathcal{J}_{\tilde{q},\tilde{s},\lambda}^{D}\right|  _{M_{\lambda}\rightarrow\infty}=\left.  \mathcal{J}_{\tilde{q},\tilde{d},\lambda}^{D}\right|  _{M_{\lambda}\rightarrow\infty } =\frac{79-8\pi^{2}}{16}\frac{M_{\tilde{q}}^{4}}{M_{\lambda}^{4}}+\mathcal{O}\left(  \frac{M_{\tilde{q}}^{6}}{M_{\lambda}^{6}}\right)  \;,\;\\
\left.  \mathcal{I}_{\tilde{q},\tilde{t},\lambda}^{M}\right|  _{M_{\lambda}\rightarrow\infty}  &  =\left.  \mathcal{J}_{\tilde{q},\tilde{s},\lambda}^{M}\right|  _{M_{\lambda}\rightarrow\infty}=\left.  \mathcal{J}_{\tilde{q},\tilde{d},\lambda}^{M}\right|  _{M_{\lambda}\rightarrow\infty } =\frac{15\pi^{2}-148}{9}\frac{M_{\tilde{q}}^{2}}{M_{\lambda}^{2}}+\frac{79-8\pi^{2}}{8}\frac{M_{\tilde{q}}^{4}}{M_{\lambda}^{4}}+\mathcal{O}\left(  \frac{M_{\tilde{q}}^{6}}{M_{\lambda}^{6}}\right)  \;.
\end{align}
Numerically, $79-8\pi^{2}\approx15\pi^{2}-148\approx1/23$, so the phase-space integrals are very suppressed when the gaugino gets much heavier than the squarks. In that case, the $\tilde{q}_{R}\rightarrow q\,t\,d\,s$ and $\tilde{q}_{L}\rightarrow q\,\bar{t}\,\bar{d}\,\bar{s}$ decay channels dominate. For our purpose, this means that same sign top quarks also arise from these four-body processes, for example via $u\,u\rightarrow\tilde{u}_{L}\tilde{u}_{L}\rightarrow\bar{t}\,\bar{t}+6j$ or $u\,u\rightarrow\tilde{u}_{R}\tilde{u}_{R}\rightarrow t\,t+6j$.

The expressions for the neutralino-induced processes $\tilde{\ell}_{L,R}\rightarrow\ell X$ and $\tilde{\nu}_{L}\rightarrow\nu X$ with $X=\bar{t}\,\bar{d}\,\bar{s}$ or $t\,d\,s$ are trivially obtained from Eq.~(\ref{Ampli4B}) and Eq.~(\ref{Rate4B}) by replacing the quark hypercharges by the adequate lepton ones, $Y_{q_{L,R}}\rightarrow Y_{\ell_{L,R}}$ and $Y_{q_{L}}\rightarrow Y_{\nu_{L}}$. In this case though, the initial state needs not have a mass close to the virtual squarks. The whole amplitude can be expanded as a series in $M_{\tilde{q},\tilde{\chi}}\rightarrow\infty$ before performing the phase-space integration, giving for $m_{t}=0$ (very similar expressions were obtained in Ref.~\cite{RPVRGE} for the $\tilde{\ell}\rightarrow\ell\,\bar{\ell
}^{\prime}\,\bar{u}\,d$ decay rate induced by the $\bm{\lambda}_{\ell^{\prime}ud}^{\prime}$ coupling)
\begin{align}
\frac{\Gamma(\tilde{\ell}_{L}\overset{}{\rightarrow}\ell\,t\,d\,s)_{\tilde{\chi
}_{1}^{0}}}{Y_{\ell_{L}}^{2}\Gamma_{\tilde{\chi}}^{0}}  &  =\frac{\Gamma
(\tilde{\ell}_{R}\overset{}{\rightarrow}\ell\,\bar{t}\,\bar{d}\,\bar{s}%
)_{\tilde{\chi}_{1}^{0}}}{Y_{\ell_{R}}^{2}\Gamma_{\tilde{\chi}}^{0}%
}=\frac{Y_{u_{R}}^{2}+2Y_{d_{R}}^{2}}{720}\frac{M_{\tilde{\ell}}^{4}%
}{M_{\tilde{\chi}}^{4}}\frac{M_{\tilde{\ell}}^{4}}{M_{\tilde{q}}^{4}}%
\times\left(  1+\frac{1}{2}\right)  \;,\smallskip\\
\frac{\Gamma(\tilde{\ell}_{R}\overset{}{\rightarrow}\ell\,t\,d\,s)_{\tilde{\chi
}_{1}^{0}}}{Y_{\ell_{R}}^{2}\Gamma_{\tilde{\chi}}^{0}}  &  =\frac{\Gamma
(\tilde{\ell}_{L}\overset{}{\rightarrow}\ell\,\bar{t}\,\bar{d}\,\bar{s}%
)_{\tilde{\chi}_{1}^{0}}}{Y_{\ell_{L}}^{2}\Gamma_{\tilde{\chi}}^{0}%
}=\frac{Y_{u_{R}}^{2}+2Y_{d_{R}}^{2}}{360}\frac{M_{\tilde{\ell}}^{2}%
}{M_{\tilde{\chi}}^{2}}\left(  1+\frac{M_{\tilde{\ell}}^{2}}{M_{\tilde{\chi}%
}^{2}}\right)  \frac{M_{\tilde{\ell}}^{4}}{M_{\tilde{q}}^{4}}\times\left(
1+\frac{1}{2}\right)  \;,
\end{align}
where the $1/2$ originate from the interference terms (as in Eq.~(\ref{RateSeries})), $Y_{u_{R}}^{2}+2Y_{d_{R}}^{2}=8/3$, $Y_{\ell_{L}}=-1$, $Y_{\ell_{R}}=2$, and $\Gamma(\tilde{\ell}_{L}\rightarrow\ell X)_{\tilde{\chi}_{1}^{0}}=\Gamma(\tilde{\nu}_{L}\rightarrow\nu X)_{\tilde
{\chi}_{1}^{0}}$ since $Y_{\nu_{L}}=Y_{\ell_{L}}$ and $Y_{\nu_{R}}=0$.


\begin{thebibliography}{99}                                                                                                

\bibitem {AtlasSUSYReach}For a compendium of the latest Atlas limits, see
\newline https://twiki.cern.ch/twiki/bin/view/AtlasPublic/SupersymmetryPublicResults.

\bibitem {CMSSUSYReach}For a compendium of the latest CMS limits, see
\newline https://twiki.cern.ch/twiki/bin/view/CMSPublic/PhysicsResultsSUS.

\bibitem {Barbier04}For a review, see e.g. R.~Barbier \textit{et al.}, Phys.\ Rept.\ \textbf{420} (2005) 1 [hep-ph/0406039].
%%CITATION = PRPLC,420,1;%%

\bibitem {PDG} J.~Beringer {\it et al.} [Particle Data Group Collaboration], Phys.\ Rev.\ D {\bf 86} (2012) 010001.
%%CITATION = PHRVA,D86,010001;%%

\bibitem {RPVMFV}E.~Nikolidakis and C.~Smith, Phys.\ Rev.\ D \textbf{77}
(2008) 015021 [arXiv:0710.3129 [hep-ph]];
%%CITATION = PHRVA,D77,015021;%%
C.~Smith, Phys.\ Rev.\ D \textbf{85} (2012) 036005 [arXiv:1105.1723 [hep-ph]].
%%CITATION = ARXIV:1105.1723;%%

\bibitem {MFV}L.~J.~Hall and L.~Randall, Phys.\ Rev.\ Lett.\ \textbf{65} (1990) 2939;
%%CITATION = PRLTA,65,2939;%%
G.~D'Ambrosio, G.~F.~Giudice, G.~Isidori and A.~Strumia, Nucl.\ Phys.\ B \textbf{645} (2002) 155.
%%CITATION = NUPHA,B645,155;%%

\bibitem {Grossman}C.~Csaki, Y.~Grossman and B.~Heidenreich, Phys.\ Rev.\ D
\textbf{85} (2012) 095009 [arXiv:1111.1239 [hep-ph]].
%%CITATION = ARXIV:1111.1239;%%

\bibitem {Gauged}G.~Krnjaic and D.~Stolarski, G.~Krnjaic and D.~Stolarski,
JHEP \textbf{1304} (2013) 064 [arXiv:1212.4860 [hep-ph]];
%%CITATION = ARXIV:1212.4860;%%
R.~Franceschini and R.~N.~Mohapatra, JHEP \textbf{1304} (2013) 098 [arXiv:1301.3637 [hep-ph]].
[arXiv:1301.3637 [hep-ph]];
%%CITATION = ARXIV:1301.3637;%%
C.~Csaki and B.~Heidenreich, arXiv:1302.0004 [hep-ph].
%%CITATION = ARXIV:1302.0004;%%

\bibitem {RPVRGE}B.~C.~Allanach, A.~Dedes and H.~K.~Dreiner, Phys.\ Rev.\ D \textbf{69} (2004) 115002 [Erratum-ibid.\ D \textbf{72} (2005) 079902] [hep-ph/0309196].
%%CITATION = HEP-PH/0309196;%%

\bibitem {DurieuxGMS12}G.~Durieux, J.-M.~G\'erard, F.~Maltoni and C.~Smith,
Phys.\ Lett.\ B \textbf{721} (2013) 82 [arXiv:1210.6598 [hep-ph]].
%%CITATION = ARXIV:1210.6598;%%

\bibitem {Allanach:2012vj}B.~C.~Allanach and B.~Gripaios, JHEP \textbf{1205} (2012) 062 [arXiv:1202.6616 [hep-ph]].
%%CITATION = ARXIV:1202.6616;%%

\bibitem {AsanoRS13}M.~Asano, K.~Rolbiecki and K.~Sakurai, JHEP \textbf{1301} (2013) 128 [arXiv:1209.5778 [hep-ph]].
%%CITATION = ARXIV:1209.5778;%%

\bibitem {BergerPST13}J.~Berger, M.~Perelstein, M.~Saelim and P.~Tanedo, JHEP \textbf{1304} (2013) 077 [arXiv:1302.2146 [hep-ph]].
%%CITATION = ARXIV:1302.2146;%%

\bibitem {Stops}D.~Choudhury, M.~Datta and M.~Maity, JHEP \textbf{1110} (2011) 004 [arXiv:1106.5114 [hep-ph]];
%%CITATION = ARXIV:1106.5114;%%
J.~A.~Evans and Y.~Kats, JHEP \textbf{1304} (2013) 028 [arXiv:1209.0764[hep-ph]];
%%CITATION = ARXIV:1209.0764;%%
Z.~Han, A.~Katz, M.~Son and B.~Tweedie, arXiv:1211.4025 [hep-ph];
%%CITATION = ARXIV:1211.4025;%%
R.~Franceschini and R.~Torre, arXiv:1212.3622 [hep-ph].
%%CITATION = ARXIV:1212.3622;%%

\bibitem {ThreeJets}S.~Chatrchyan \textit{et al.} [CMS Collaboration],
Phys.\ Lett.\ B \textbf{718} (2012) 329 [arXiv:1208.2931 [hep-ex]];
%%CITATION = ARXIV:1208.2931;%%
G.~Aad \textit{et al.} [ATLAS Collaboration], JHEP \textbf{1212} (2012) 086 [arXiv:1210.4813 [hep-ex]].
%%CITATION = ARXIV:1210.4813;%%

\bibitem {Rhadrons}M.~Fairbairn, A.~C.~Kraan, D.~A.~Milstead, T.~Sjostrand,
P.~Z.~Skands and T.~Sloan, Phys.\ Rept.\ \textbf{438} (2007) 1 [hep-ph/0611040].
%%CITATION = HEP-PH/0611040;%%

\bibitem {StablesQ}See [ATLAS Collaboration], ATLAS-CONF-2012-075, and references there.
%%CITATION = ATLAS-CONF-2012-075;%%

\bibitem {SplitSUSY}N.~Arkani-Hamed and S.~Dimopoulos, JHEP \textbf{0506} (2005) 073 [hep-th/0405159];
%%CITATION = HEP-TH/0405159;%%
G.~F.~Giudice and A.~Romanino, Nucl.\ Phys.\ B \textbf{699} (2004) 65 [Erratum-ibid.\ B \textbf{706} (2005) 65] [hep-ph/0406088].
%%CITATION = HEP-PH/0406088;%%

\bibitem {Buchkremer:2012dn} M.~Buchkremer and A.~Schmidt, Adv.\ High Energy Phys.\ \textbf{2013}, 690254 (2013) [arXiv:1210.6369 [hep-ph]].
%%CITATION = ARXIV:1210.6369;%%

\bibitem {Monotops}J.~Andrea, B.~Fuks and F.~Maltoni, Phys.\ Rev.\ D \textbf{84} (2011) 074025 [arXiv:1106.6199 [hep-ph]].
%%CITATION = ARXIV:1106.6199;%%

\bibitem {MFVRGE}G.~Colangelo, E.~Nikolidakis and C.~Smith, Eur.\ Phys.\ J.\ C {\bf 59} (2009) 75
[arXiv:0807.0801 [hep-ph]].
%%CITATION = ARXIV:0807.0801;%%

\bibitem {MaxMixetal}See A.~Arbey, M.~Battaglia, A.~Djouadi, F.~Mahmoudi and J.~Quevillon, Phys.\ Lett.\ B \textbf{708} (2012) 162 [arXiv:1112.3028 [hep-ph]], and references there.
%%CITATION = ARXIV:1112.3028;%%

\bibitem {Berdine07}D.~Berdine, N.~Kauer and D.~Rainwater, Phys.\ Rev.\ Lett.\ \textbf{99} (2007) 111601 [hep-ph/0703058].
%%CITATION = HEP-PH/0703058;%%

\bibitem {DiracGlue} S.~Y.~Choi, M.~Drees, A.~Freitas and P.~M.~Zerwas, Phys.\ Rev.\ D \textbf{78} (2008) 095007  [arXiv:0808.2410 [hep-ph]].  
%%CITATION = ARXIV:0808.2410;%%

\bibitem {Christensen:2008py}N.~D.~Christensen and C.~Duhr, Comput.\ Phys.\ Commun.\ \textbf{180} (2009) 1614 [arXiv:0806.4194 [hep-ph]].
%%CITATION = ARXIV:0806.4194;%%

\bibitem {Alwall:2011uj}J.~Alwall, M.~Herquet, F.~Maltoni, O.~Mattelaer and T.~Stelzer, JHEP \textbf{1106} (2011) 128 [arXiv:1106.0522 [hep-ph]].
%%CITATION = ARXIV:1106.0522;%%

\bibitem {ggNLO} See for example M.~Kramer, A.~Kulesza, R.~van der Leeuw, M.~Mangano, S.~Padhi, T.~Plehn and X.~Portell, arXiv:1206.2892 [hep-ph], and the numerical estimates at the 8~TeV LHC published on-line at https://twiki.cern.ch/twiki/bin/view/LHCPhysics/SUSYCrossSections.
%%CITATION = ARXIV:1206.2892;%%

\bibitem {CMS7}S.~Chatrchyan \textit{et al.} [CMS Collaboration], JHEP \textbf{1208} (2012) 110 [arXiv:1205.3933 [hep-ex]].
%%CITATION = ARXIV:1205.3933;%%

\bibitem {CMS8}S.~Chatrchyan \textit{et al.} [CMS Collaboration], JHEP \textbf{1303} (2013) 037 [arXiv:1212.6194 [hep-ex]];
%%CITATION = ARXIV:1212.6194;%%
CMS PAS SUS-12-017; CMS PAS SUS-12-029; S.~Chatrchyan \textit{et al.} [CMS Collaboration], Phys.\ Rev.\ Lett.\ \textbf{109} (2012) 071803 [arXiv:1205.6615 [hep-ex]].
%%CITATION = ARXIV:1205.6615;%%

\bibitem {Atlas7}G.~Aad \textit{et al.} [ATLAS Collaboration], Phys.\ Rev.\ Lett.\ \textbf{108} (2012) 241802 [arXiv:1203.5763 [hep-ex]];
%%CITATION = ARXIV:1203.5763;%%
JHEP \textbf{1212} (2012) 007 [arXiv:1210.4538 [hep-ex]].
%%CITATION = ARXIV:1210.4538;%%

\bibitem {Atlas8}G.~Aad \textit{et al.} [ATLAS Collaboration], ATLAS-CONF-2012-105; ATLAS-CONF-2013-007; ATLAS-CONF-2013-051.

\bibitem {SMttW}J.~M.~Campbell and R.~K.~Ellis, JHEP \textbf{1207} (2012) 052 [arXiv:1204.5678 [hep-ph]].
%%CITATION = ARXIV:1204.5678;%%

\bibitem {SMttZ}A.~Kardos, Z.~Trocsanyi and C.~Papadopoulos, Phys.\ Rev.\ D \textbf{85} (2012) 054015 [arXiv:1111.0610 [hep-ph]];
%%CITATION = ARXIV:1111.0610;%%
M.~V.~Garzelli, A.~Kardos, C.~G.~Papadopoulos and Z.~Trocsanyi, Phys.\ Rev.\ D \textbf{85} (2012) 074022 [arXiv:1111.1444 [hep-ph]].
%%CITATION = ARXIV:1111.1444;%%

\bibitem {Boosted} J.~Thaler and L.~-T.~Wang, JHEP \textbf{0807} (2008) 092 [arXiv:0806.0023 [hep-ph]];
%%CITATION = ARXIV:0806.0023;%%
K.~Rehermann and B.~Tweedie, JHEP \textbf{1103} (2011) 059 [arXiv:1007.2221 [hep-ph]];
%%CITATION = ARXIV:1007.2221;%%
T.~Plehn, M.~Spannowsky and M.~Takeuchi, JHEP \textbf{1105} (2011) 135 [arXiv:1102.0557 [hep-ph]].
%%CITATION = ARXIV:1102.0557;%%

\bibitem {BaltzG97}E.~A.~Baltz and P.~Gondolo, Phys.\ Rev.\ D \textbf{57} (1998) 2969 [hep-ph/9709445].
%%CITATION = HEP-PH/9709445;%%

\bibitem {ButterworthD92}J.~Butterworth and H.~K.~Dreiner, Nucl.\ Phys.\ B \textbf{397} (1993) 3 [hep-ph/9211204].
%%CITATION = HEP-PH/9211204;%%

\bibitem {DreinerRS99}H.~K.~Dreiner, P.~Richardson and M.~H.~Seymour, JHEP \textbf{0004} (2000) 008 [hep-ph/9912407].
%%CITATION = HEP-PH/9912407;%%

\bibitem {Kramer09}M.~Kramer, E.~Popenda, M.~Spira and P.~M.~Zerwas, Phys.\ Rev.\ D \textbf{80} (2009) 055002 [arXiv:0902.3795 [hep-ph]].
%%CITATION = ARXIV:0902.3795;%%

\end{thebibliography}
\end{document}